\documentclass[12pt]{article}
\usepackage{color}
\usepackage{latexsym}
\usepackage{epsfig,amssymb,euscript, mathrsfs,cite}
\usepackage{amsmath}
\definecolor{MyDarkBlue}{rgb}{0.15,0.15,0.45}
\usepackage[linktocpage=true]{hyperref}
\usepackage{enumerate}
\hypersetup{
colorlinks=true,
citecolor=MyDarkBlue,
linkcolor=MyDarkBlue,
urlcolor=MyDarkBlue,
pdfauthor={},
pdftitle={},
pdfsubject={hep-th}
}
\newsavebox{\ns}
\newsavebox{\dbrane}
\newsavebox{\dbshort}

\def\be{\begin{equation}}
\def\ee{\end{equation}}
\def\bea{\begin{eqnarray}}
\def\eea{\end{eqnarray}}

\newcommand{\nn}{\nonumber}

\newcommand\R{\mathbb{R}}
\newcommand\Z{\mathbb{Z}}

\newcommand\C{\mathbb{C}}

\newcommand\diff{\mathrm{d}}

\newcommand{\dd}{\mathrm{d}}

\newcommand{\ii}{\mathrm{i}}

\newcommand{\ex}{\mathrm{e}}
\newcommand{\vol}{\mathrm{vol}}

\newlength{\sswidth}

\newcommand{\Cweyl}{\mathscr{C}}
\newcommand{\hook}{\mathbin{\rule[.2ex]{.4em}{.03em}\rule[.2ex]{.03em}{.9ex}}}

\numberwithin{equation}{section}       % equation numbers in each section

\begin{document}

\bibliographystyle{utphys}

\begin{titlepage}

\begin{center}

\today

\vskip 2.3 cm 

\vskip 5mm

{\Large \bf Supersymmetric gauge theories on five-manifolds}

\vskip 15mm

{Luis F. Alday, Pietro Benetti Genolini, Martin Fluder,}\\ \vskip 3mm
{ Paul Richmond and James Sparks}

\vspace{1cm}
\centerline{{\it Mathematical Institute, University of Oxford,}}
\centerline{{\it Andrew Wiles Building, Radcliffe Observatory Quarter,}}
\centerline{{\it Woodstock Road, Oxford, OX2 6GG, UK}}

\end{center}

\vskip 2 cm

\begin{abstract}
\noindent We construct rigid supersymmetric gauge theories on Riemannian five-manifolds. 
We follow a holographic approach, realizing the manifold as the conformal boundary of a six-dimensional bulk supergravity solution. This leads to a systematic classification of five-dimensional supersymmetric backgrounds with gravity duals.  
We show that the background metric is furnished with a conformal Killing vector, which generates a transversely holomorphic foliation with a transverse Hermitian structure. Moreover, we prove that any such metric defines a supersymmetric background.
Finally, we construct supersymmetric Lagrangians for gauge theories coupled to arbitrary matter on such backgrounds.
\end{abstract}

\end{titlepage}

\pagestyle{plain}
\setcounter{page}{1}
\newcounter{bean}
\baselineskip18pt
\tableofcontents

%%%%%%%%%%%%%%%%%%%%%%%%%%%%%%%%%%%%%%%%%%%

\section{Introduction}\label{SecIntroduction}

There has recently been considerable work on defining and studying supersymmetric gauge theories on curved backgrounds. 
The main reason for this interest is that these quantum field theories
 possess classes of observables that may be computed exactly using localization methods. 
 Such non-perturbative results allow for quantitative tests of various conjectured dualities, and 
 have also led to the discovery of new dualities. A primary example is the AdS/CFT correspondence, 
 where exact strong-coupling field theory calculations may be compared to semi-classical gravity. 

In this paper we focus on rigid supersymmetry in $d=5$ dimensions, which is currently not as 
well-developed as its lower-dimensional cousins. Supersymmetric gauge theories 
were constructed and studied on the round $S^5$ in \cite{Kallen:2012cs, Hosomichi:2012ek, Kallen:2012va, Kim:2012ava}. 
The product background $S^1\times S^4$ studied in \cite{Kim:2012gu, Terashima:2012ra} leads 
to the superconformal index. As in lower dimensions, the first constructions 
of non-conformally flat backgrounds were produced via various \emph{ad hoc} methods. 
These include the squashed $S^5$ geometries of \cite{Imamura:2012xg, Imamura:2012bm}, 
and the product backgrounds $S^3\times \Sigma_2$ \cite{Kawano:2012up, Fukuda:2012jr} 
and $S^2\times M_3$ \cite{Yagi:2013fda, Lee:2013ida, Cordova:2013cea}. 
In the latter two cases the spheres are round, while supersymmetry on the 
Riemann surface $\Sigma_2$ or three-manifold $M_3$ is achieved via a topological twist 
utilizing the $SU(2)_R$ symmetry of the theory. These constructions have been used to 
successfully test AGT-type correspondences. 

A systematic method for constructing rigid supersymmetric field theories 
on curved backgrounds, in any dimension $d$,
 was initiated in \cite{Festuccia:2011ws}. Here one first couples 
 the field theory to off-shell supergravity, and then takes a decoupling 
 limit in which the gravity multiplet becomes a non-dynamical background field. 
This approach was applied to five-dimensional Poincar\'e supergravity \cite{Zucker:1999ej, Kugo:2000hn, Kugo:2000af} 
in the series of papers \cite{Pan:2013uoa, Imamura:2014ima, Pan:2014bwa}.\footnote{See \cite{Kuzenko:2014eqa} for the construction of supersymmetric Lorentzian backgrounds within the superspace formulation of five-dimensional conformal supergravity.} 
Supersymmetry of the background requires a certain generalized Killing 
spinor equation to hold, whose related geometry was investigated in \cite{Pan:2013uoa}, 
together with an algebraic ``dilatino'' equation which was studied in \cite{Imamura:2014ima}. 
The latter reference recasts these conditions into local geometric constraints 
on the five-manifold $M_5$.  As in lower dimensions, one finds that the background 
is parametrized by various arbitrary functions/tensors. In particular $(M_5,g)$ is 
equipped with a Killing vector field $\xi=\partial_\psi$, with dual one-form $S^2(\diff\psi+\rho)$ 
and transverse four-dimensional metric $g^{(4)}$, where locally the function $S=\|\xi\|$ and tensors $\rho$ and $g^{(4)}$ 
are $\xi$-invariant but otherwise freely specifiable.\footnote{There are also additional freely specifiable fields,
which determine the rest of the 
background.} The authors of \cite{Imamura:2014ima}
furthermore show that \emph{locally} all deformations of the background fields 
lead to $Q$-exact deformations of the action, where $Q$ is the supercharge. 
Despite this generality, these backgrounds 
apparently don't include  the conformally flat $S^1\times S^4$ geometry mentioned above \cite{Imamura:2014ima}. 
We shall 
comment further on these issues later.

%An entirely different approach to defining supersymmetry on five-manifolds is followed in \cite{ Kallen:2012cs,  Qiu:2014cha}. 
In 
\cite{Kallen:2012cs} a twisted version of $\mathcal{N}=1$ super-Yang-Mills theory 
is defined on \emph{contact} five-manifolds $(M_5,\eta)$. Here $\eta$ is a contact 
one-form, meaning that $\eta\wedge \diff\eta\wedge \diff\eta$ is a volume form. 
On a Sasaki-Einstein five-manifold \cite{Sparks:2010sn} one can construct $\mathcal{N}=1$ 
super-Yang-Mills coupled to matter \cite{Qiu:2013pta}. This is essentially because the 
two Killing spinors on a Sasaki-Einstein manifold satisfy the same 
Killing spinor equations as those on the round sphere. 
For the special class of toric ($U(1)^3$-invariant)
Sasaki-Einstein manifolds of \cite{Martelli:2005tp}
the localized perturbative partition function has been computed in
\cite{Qiu:2013aga, Schmude:2014lfa, Qiu:2014oqa}, with the 
last reference also giving a conjectured formula for the full partition 
function. The authors of \cite{Qiu:2014cha} furthermore show that 
one can define a twisted version of $\mathcal{N}=2$ super-Yang-Mills theory  
on any \emph{K-contact} five-manifold. 
We also note that K-contact 
geometry arises as a special case 
in \cite{Pan:2013uoa}.

In the present paper we instead take a holographic approach, similar to \cite{Klare:2012gn} in lower dimensions, 
 to construct
rigid supersymmetry in five dimensions. 
Here $M_5$ is realized as the conformal boundary of a six-dimensional bulk solution of  Romans $F(4)$ gauged supergravity \cite{Romans:1985tw}. 
Some of the groundwork for this was laid in \cite{Alday:2014rxa, Alday:2014bta}, 
where supergravity duals of the squashed five-sphere backgrounds of \cite{Imamura:2012xg, Imamura:2012bm} 
were constructed (see also \cite{Alday:2014fsa, Hama:2014iea} for holographic duals 
to the supersymmetric R\'enyi entropy in five dimensions). We begin with a 
general supersymmetric asymptotically locally AdS solution to the Romans 
theory, and extract the conditions this imposes on the five-dimensional conformal boundary. 
Although the resulting spinor equations are quite complicated, we will show they 
are completely equivalent to a very simple geometric structure.  We find that $M_5$ is 
equipped with a conformal Killing vector $\xi=\partial_\psi$ which generates 
a \emph{transversely holomorphic foliation}. This is compatible with 
an almost contact form $\eta=\diff\psi+\rho$, where up to global 
constraints that we describe the norm $S=\|\xi\|$ and $\rho$ are arbitrary, 
and the transverse metric $g^{(4)}$ is \emph{Hermitian}. The only other remaining 
freedom is an arbitrary function $\alpha$ (such that $S\alpha$ is $\xi$-invariant), which together with the metric 
determines all the remaining background data. This structure is similar to the rigid limit of Poincar\'e supergravity described above, 
but with the addition of an integrable transverse complex structure and Hermitian metric. 
In fact it is a natural hybrid of the ``real'' three-dimensional rigid supersymmetric 
geometry studied in \cite{Closset:2012ru, Alday:2013lba} and the four-dimensional 
supersymmetric geometry of \cite{Dumitrescu:2012ha, Klare:2013dka} (where 
the four-manifold is complex with a compatible Hermitian metric). 

The outline of the rest of the paper is as follows. In section \ref{SecHol} 
we summarize the form of  supersymmetric asymptotically locally AdS solutions to Romans
supergravity, in particular extracting the Killing spinor 
equations on the conformal boundary $M_5$. These are then used 
as a starting point for a purely five-dimensional analysis
in section~\ref{SecGeometry}. We show that the spinor equations 
are completely equivalent to
 a simple geometric structure on $M_5$, and present a number of 
subclasses and examples, including many of the examples referred to above. In section \ref{SecLagrangians} we construct $\mathcal{N}=1$ 
supersymmetric gauge theories formed of vector
and hypermultiplets on this background geometry. Our 
conclusions are presented in section \ref{SecDiscussion}.

%%%%%%%%%%%%%%%%%%%%%%%%%%%%%%%%%%%%%%%%%%%

\section{Rigid supersymmetry from holography}\label{SecHol}

The bosonic fields of the six-dimensional Romans supergravity theory \cite{Romans:1985tw} consist of the metric, a scalar field $X$, a two-form potential $B$, together with an $SO(3)_R\sim SU(2)_R$ R-symmetry gauge field $A^i$ with field strength $F^i=\diff A^i-\frac{1}{2}\varepsilon_{ijk}A^j\wedge A^k$, where $i=1,2,3$. 
Here we are working in a gauge in which the Stueckelberg one-form is zero, and we set 
the  gauge coupling constant to 1. 
  The Euclidean signature equations of motion for this theory may be found 
 in \cite{Alday:2014bta}, although we will not require their explicit form here.

A  solution is 
 supersymmetric provided there 
exists a non-trivial $SU(2)_R$ doublet of Dirac spinors $\epsilon_I$, $I=1,2$, satisfying 
the following Killing spinor  and dilatino equations
\bea
D_M \epsilon_I & =&  \frac{\ii}{4\sqrt{2}}  ( X + \tfrac{1}{3} X^{-3} ) \Gamma_M \Gamma_7 \epsilon_I - \frac{\ii}{16\sqrt{2}} X^{-1} F_{NP} ( \Gamma_M{}^{NP} - 6 \delta_M{}^N \Gamma^P ) \epsilon_I \label{KSE}  \\
&& - \frac{1}{48} X^2 H_{NPQ} \Gamma^{NPQ} \Gamma_M \Gamma_7 \epsilon_I
+ \frac{1}{16\sqrt{2}}X^{-1} F_{NP}^i ( \Gamma_M{}^{NP} - 6 \delta_M{}^N \Gamma^P) \Gamma_7 ( \sigma^i )_I{}^J \epsilon_J ~,\nn\\
0 & = & - \ii X^{-1} \partial_M X \Gamma^M \epsilon_ I + \frac{1}{2\sqrt{2}}  \left( X - X^{-3} \right) \Gamma_7 \epsilon_I + \frac{\ii}{24} X^2 H_{MNP} \Gamma^{MNP} \Gamma_7 \epsilon_I \nonumber \\
&&- \frac{1}{8\sqrt{2}} X^{-1} F_{MN} \Gamma^{MN} \epsilon_I - \frac{\ii}{8\sqrt{2}} X^{-1} F^i_{MN} \Gamma^{MN} \Gamma_7 ( \sigma^i )_I{}^J \epsilon_J~.\label{dilatino}
\eea
Here  $\Gamma_M$ are taken to be Hermitian and generate the Clifford 
algebra $\mathrm{Cliff}(6,0)$ in an orthonormal frame, $M=0,\ldots,5$. We have defined the chirality operator
 $\Gamma_7 = \ii \Gamma_{012345}$, which satisfies $(\Gamma_7)^2=1$. 
The covariant derivative acting on the spinor is $D_M\epsilon_I=\hat{\nabla}_M\epsilon_I+\frac{\ii}{2} A^i_M(\sigma_i)_I{}^J\epsilon_J$, where $\hat{\nabla}_M=\partial_M+\frac{1}{4}\Omega_{M}^{\ \ NP}\Gamma_{NP}$ denotes the Levi-Civita 
spin connection while $\sigma_i$, $i=1,2,3$, are the Pauli matrices.

Given a supersymmetric asymptotically locally AdS solution we may 
introduce a radial coordinate $r$, so that the conformal boundary is at $r=\infty$ and 
the metric admits an expansion of the form
\bea\label{diagonalmetric}
\diff s^2 &=& \frac{9}{2}\frac{\diff r^2}{r^2}+ r^2\left[g_{\mu\nu}+\frac{1}{r^2}g^{(2)}_{\mu\nu} + \cdots\right]\diff x^\mu\diff x^\nu~.
\eea
Here $x^\mu$, $\mu=1,\ldots,5$, are coordinates on the conformal boundary, which has 
metric $g=(g_{\mu\nu})$.  Notice that the particular form of the metric in (\ref{diagonalmetric}) is 
not reparametrization invariant under $r\rightarrow \Lambda r$, where $\Lambda=\Lambda(x^\mu)$. 
However, the correction terms under such a transformation are subleading in the $1/r$ expansion. 
This will play an important role in the next section. 

For simplicity we shall mainly consider Abelian solutions in which 
$A^1=A^2=0$, and $A^3\equiv A$, with field strength $F\equiv \diff A$. 
Similarly to the metric (\ref{diagonalmetric})
we then write the following general expansions for the remaining bosonic fields
\bea\label{expand}
X &=& 1 + \frac{1}{r^2}X_2+\cdots~,\nn\\
B &=& r b - \frac{1}{r^2}\diff r\wedge A^{(0)}+\cdots\nn\\
A &=& a  +  \cdots~,
\eea
where we define $f\equiv \diff a$.
Some of the terms {\it a priori} present  in these expansions are set to zero by the equations of motion; 
for example, the $O(1/r$) term in the expansion of $X$ \cite{Alday:2014bta}. 
The Killing spinors similarly admit an expansion of the form
\bea\label{epsilonFernando}
\epsilon_I &=& \sqrt{r}\begin{pmatrix}\chi_I \\ -\ii \chi_I\end{pmatrix}+ \frac{1}{\sqrt{r}}\begin{pmatrix}
\varphi_I \\ \ii \varphi_I\end{pmatrix} + O(r^{-3/2})~.
\eea
Here we have used the orthonormal frame
\bea
E^0 &=& \frac{3}{\sqrt{2}}\frac{\diff r}{r}~, \qquad E^\mu \ = \ r e^\mu + \cdots
\eea
for the metric (\ref{diagonalmetric}). Furthermore, the spin connection expands as
\bea
\Omega_\mu^{\ \, 0\nu} &=& -\frac{\sqrt{2}}{3}\delta_{\mu}^{\ \nu} + \frac{1}{r^2}\omega_\mu^{\ \nu} + \cdots~.
\eea

 Also as in \cite{Alday:2014bta} we consider a ``real'' class of solutions
 for which $\epsilon_I$ satisfies the symplectic Majorana condition 
 $\varepsilon_{I}^{\ J}\epsilon_J = \mathcal{C}_6\epsilon_I^*\equiv \epsilon_I^c$, where 
 $\mathcal{C}_6$ denotes the charge conjugation matrix, satisfying $\Gamma_M^T=\mathcal{C}_6^{-1}\Gamma_M \mathcal{C}_6$. The bosonic fields 
 are all taken to be real, with the exception of the $B$-field which is purely imaginary. 
 With these reality properties, one can show that the Killing spinor equation (\ref{KSE}) and dilatino equation 
 (\ref{dilatino}) for $\epsilon_2$  are simply the charge conjugates of the corresponding 
 equations for $\epsilon_1$.
  In this way 
 we  effectively reduce to a single Killing spinor $\epsilon\equiv \epsilon_1$, with $SU(2)_R$ doublet $(\epsilon,\epsilon^c)$. 
 We then note the following large $r$ expansions of bilinears:\footnote{Here we take 
 the spinors to be Grassmann even.}
\bea\label{bulkbil}
\epsilon^\dagger\Gamma_7\epsilon &=& 4\alpha S + \cdots~,\nn\\
\ii \epsilon^\dagger \Gamma_7\Gamma_{(1)}\epsilon &=& 2S r K_2 - 3\sqrt{2}\diff r + \cdots~.
\eea
Here we have defined $\Gamma_{(1)}\equiv \Gamma_M E^M$ and
\bea
S &\equiv & \chi^\dagger\chi~.
\eea
We also note that the bilinear $\epsilon^\dagger\Gamma_{(1)}\epsilon$ is a Killing one-form in the bulk \cite{Alday:2014bta}. 
This will hence restrict to a conformal Killing vector on the boundary at $r=\infty$. 

Substituting the expansions (\ref{epsilonFernando}) into the bulk Killing spinor equation (\ref{KSE}), at the first two orders we obtain
\bea\label{KSEchi}
\left(\nabla_\mu +\frac{\ii}{2}a_\mu\right)\chi &=& -\frac{\sqrt{2}}{3}\ii\gamma_\mu \varphi 
- \frac{\ii}{12\sqrt{2}}b_{\nu\sigma}\gamma_\mu^{\ \nu\sigma}\chi + \frac{\ii}{3\sqrt{2}}b_{\mu\nu}\gamma^\nu\chi~,\\\label{KSEvarphi}
\left(\nabla_\mu  + \frac{\ii}{2}a_\mu\right)\varphi &=& -\frac{\ii}{6\sqrt{2}}b_{\mu\nu}\gamma^{\nu}\varphi 
+ \frac{1}{16\sqrt{2}}f_{\nu\sigma}\gamma_{\mu}^{\ \nu\sigma}\chi - \frac{3}{8\sqrt{2}}f_{\mu\nu}\gamma^\nu\chi\\
&& + \frac{1}{48}(\diff b)_{\nu\rho\sigma}\gamma^{\nu\rho\sigma}\gamma_\mu\chi - \frac{1}{36}A^{(0)}_\nu\gamma_\mu^{\ \nu}\chi + \frac{1}{12}A^{(0)}_\mu\chi + \frac{\ii}{2}\omega_\mu^{\ \nu}\gamma_\nu \chi~.\nn\eea
Here $\gamma_\mu$ generate the Clifford algebra $\mathrm{Cliff}(5,0)$ in an orthonormal frame,  while
$\nabla$ denotes the Levi-Civita spin connection for the boundary metric $g$. 
Similarly, the bulk dilatino equation (\ref{dilatino}) implies
\bea\label{Dilchi}
-\frac{1}{6\sqrt{2}}b_{\mu\nu}\gamma^{\mu\nu}\varphi -\frac{\sqrt{2}}{3}X_2\chi + \frac{\ii}{8\sqrt{2}}f_{\mu\nu}\gamma^{\mu\nu}\chi+\frac{\ii}{24}(\diff b)_{\mu\nu\sigma}\gamma^{\mu\nu\sigma}\chi 
-\frac{\ii}{18}A^{(0)}_\mu\gamma^\mu \chi \, = \, 0~.
\eea

As explained in \cite{Alday:2014bta}, equation (\ref{KSEchi}) may be rewritten in the form of a charged 
conformal Killing spinor equation, with additional $b$-field couplings. Setting $b=0$ one 
obtains the standard charged conformal Killing spinor equation, whose solutions (twistor spinors)
have been  studied in the holographic context for three-manifolds and four-manifolds in 
\cite{Klare:2012gn, Klare:2013dka, Hristov:2013spa, Cassani:2012ri, Lischewski}. 
On the other hand, previous work  on rigid supersymmetry in five dimensions \cite{Pan:2013uoa, Imamura:2014ima, Pan:2014bwa} has used 
Killing spinor equations of a different form, without the coupling to $\varphi$ in (\ref{KSEchi}). We may make closer  contact with this work 
by noting that supersymmetry in the bulk also implies the algebraic relation
\bea\label{Dilvarphi}
\varphi &=& -\alpha\chi - \frac{\ii}{2}(K_2)_\nu\gamma^\nu\chi~.
\eea
This follows from the bilinear expansions (\ref{bulkbil}).

In the remainder of the paper we shall take equations  (\ref{KSEchi}), (\ref{KSEvarphi}), 
(\ref{Dilchi}), and (\ref{Dilvarphi})  as our starting point for a purely five-dimensional analysis. 

%%%%%%%%%%%%%%%%%%%%%%%%%%%%%%%%%%%%%%%%%%%

\section{Background geometry}\label{SecGeometry}

In this section we begin with a Riemannian five-manifold $(M_5,g)$, 
on which we'd like to define rigid supersymmetric gauge theories. The 
gauge/gravity correspondence implies this should be possible, provided 
the spinor equations derived in the previous section hold. 

Let us summarize the background data. In addition to the real metric 
$g$, we have two generalized Killing spinors $\chi$, $\varphi$.  Globally these are spin$^c$ spinors, 
being sections of the spin bundle of $M_5$ tensored with $L^{-1/2}$, $\chi,\varphi \in \Gamma[\mathrm{Spin}(M_5)\otimes L^{-1/2}]$,
where $L$ is the complex line bundle for which the real gauge field $a$ is a connection. 
This Abelian gauge field is a background field for  $U(1)_R\subset SU(2)_R$, 
with $(\chi,\chi^c)$, $(\varphi,\varphi^c)$ forming $SU(2)_R$ doublets, where $\chi^c\equiv \mathcal{C}\chi^*$
with $\mathcal{C}$ the five-dimensional charge conjugation matrix. 
The spinors $\chi$, $\varphi$ then satisfy the coupled Killing spinor equations
(\ref{KSEchi}), (\ref{KSEvarphi}), where the background $b$-field is taken to be a purely imaginary two-form, 
$A^{(0)}$ is a purely imaginary one-form, while $\omega_{\mu\nu}=g_{\nu\sigma}\omega_{\mu}^{\ \sigma}$ is real and symmetric. 
Furthermore, $\chi$ and $\varphi$ are related algebraically by (\ref{Dilvarphi}), which introduces 
the additional background fields $\alpha$ and $K_2$, which 
are respectively a real function and real one-form. Finally we have the  
dilatino equation (\ref{Dilchi}), which introduces the real background function $X_2$.

In the remainder of this section we shall analyse the geometric constraints  that these equations impose on $(M_5,g)$. 
Although the background data and equations  (\ref{KSEchi})--(\ref{Dilvarphi})   appear {\it a priori}
complicated, in fact we shall see that the geometry they are equivalent to is very simple.

\subsection{Differential constraints}\label{SecDiffConstraints}

In the analysis that follows it is convenient to assume that the spin$^c$ spinor $\chi$ is nowhere zero. 
More generally $\chi$ could vanish along some locus $Z\subset M_5$, 
and the local geometry we shall derive below is valid on $M_5\setminus Z$. If  $Z$ is non-empty one 
would need to impose suitable boundary conditions, although we shall not consider 
this further in this paper. 
 A nowhere zero 
spin$^c$ spinor equips $(M_5,g)$ with 
a local $SU(2)$ structure. Specifically, we may define the bilinears
\bea\label{bilinears}
S & \equiv & \chi^\dagger\chi~, \qquad \qquad \ \  K_1 \ \equiv \ \frac{1}{S}\chi^\dagger\gamma_{(1)}\chi~,\nn\\
J & \equiv & -\frac{\ii}{S}\chi^\dagger\gamma_{(2)}\chi~, \qquad \Omega \ \equiv \ -\frac{1}{S}(\chi^c)^\dagger\gamma_{(2)}\chi~.
\eea
Here we have introduced the notation $\gamma_{(n)}\equiv \frac{1}{n!}\gamma_{\mu_1\cdots \mu_n}\diff x^{\mu_1}\wedge \cdots \wedge \diff x^{\mu_n}$, 
where $x^\mu$, $\mu=1,\ldots,5$, are local coordinates on $M_5$. Since 
$\chi$ is nowhere zero  the scalar function $S$ is strictly positive, and it makes sense 
to normalize the bilinears as in (\ref{bilinears}). We note that $K_1$ is a real unit length one-form, 
while $J$ is a real two-form with square length $\|J\|^2=2$. Here the square norm of a $p$-form 
$\phi$ is defined via $\|\phi\|^2\, \vol_5 = \phi\wedge *\phi$, where~$*$ denotes the Hodge duality operator on $(M_5,g)$ and $\vol_5$ denotes the Riemannian volume form. The complex bilinear $\Omega$ 
is globally a two-form valued in the line bundle $L^{-1}$. 

That $\chi$, or equivalently the bilinears (\ref{bilinears}), defines a local $SU(2)$ structure follows from 
some simple group theory. The spin group is Spin$(5)\cong Sp(2)\subset U(4)$, with the 
latter acting in the fundamental representation on the spinor space $\C^4$. The stabilizer 
of a non-zero spinor is then $Sp(1)\cong SU(2)$. When $M_5$ is spin and $L$ is trivial, so that $\chi\in \Gamma[\mathrm{Spin}(M_5)]$, this defines a 
global $SU(2)$ structure. However, more generally we require only that $M_5$ is 
spin$^c$, and in this case  the \emph{global} stabilizer 
group is enlarged to $U(2)$: the additional $U(1)$ factor rotates the spinor by a phase, 
which may  be undone by a $U(1)$ gauge transformation. 
To see this in more detail we
introduce a local orthonormal frame 
$e^a$, $a=1,\ldots,5$, so that
\bea
K_1 & =& e^5~, \qquad  
J\ =\ e^1\wedge e^2+e^{3}\wedge e^4~, \qquad \Omega \ = \ (e^1+\ii e^2)\wedge (e^3+\ii e^4)~,
\eea
where the metric is $g=\sum_{a=1}^5 (e^a)^2$. The $U(2)=SU(2)\times_{\Z_2}U(1)$ structure group acts 
in the obvious way on the $\C^2$ spanned by
$e^1+\ii e^2,e^3+\ii e^4$. This leaves $K_1$, $J$ and the metric $g$ invariant, but 
rotates $\Omega$ by the determinant of the $U(2)$ transformation. In order 
for this to be undone by a gauge transformation, this identifies the line bundle as $L=\Lambda^{2,0}$. 
The latter is the space of Hodge type $(2,0)$-forms for the four-dimensional 
vector bundle spanned by $e^1$, $e^2$, $e^3$, $e^4$, and with almost complex structure 
$I$ for which $e^1+\ii e^2$ and $e^3+\ii e^4$ are $(1,0)$-forms. 
Thus our  rigid supersymmetric geometry will in general be equipped with 
a global $U(2)$ structure on $M_5$ (or more precisely on $M_5\setminus Z$).

The one-form $SK_1=\chi^\dagger \gamma_{(1)}\chi$ arises simply from the restriction of the bulk Killing one-form $\epsilon^\dagger\Gamma_{(1)}\epsilon$ 
to the conformal boundary, and thus defines a conformal Killing one-form on $(M_5,g)$. 
This is easily confirmed from the Killing spinor equation (\ref{KSEchi}) for $\chi$, which implies
\bea
\nabla_{(\mu}(SK_1)_{\nu)} &=& \mathcal{L}_\xi (\log S) g_{\mu\nu}~,
\eea
where we have introduced the  dual vector field $\xi$, 
defined by $g(\xi,\cdot ) = SK_1$, and $\mathcal{L}$ denotes the Lie derivative. 

One  finds that the spinor equations (\ref{KSEchi})--(\ref{Dilvarphi})  imply the following differential constraints:
\bea\label{scalarbil}
\dd  S & = & \ - \frac{\sqrt{2}}{3}  \left( S K_2 + \ii i_{\xi} b \right) \, , \qquad
 \dd (S \alpha ) \  = \ \ - \frac{1}{2\sqrt{2}} i_{\xi} \dd a \, , \\
\label{K1bil}
 \dd ( S K_1 ) & = &  \frac{2\sqrt{2}}{3} \left[ 2 \alpha S J +  SK_1 \wedge K_2 + \ii  S b - \frac{\ii}{2} i_{\xi} ( * b ) \right] \, , \\
\label{K2bil}
  \dd ( S K_2 ) & = & \ii i_{\xi} \dd b -\ii \mathcal{L}_\xi ( \log S)b\, ,\\
\label{Jbil}
\dd ( S J ) & =& \ - \sqrt{2} K_2 \wedge ( S J ) \, , \\
\label{Omegabil}
\dd ( S \Omega ) & = & \ - \ii \left( a - 2 \sqrt{2} \alpha K_1 - \ii \sqrt{2} K_2 \right) \wedge ( S \Omega ) \, .
\eea
Here $(i_{V}\phi)_{a_1\cdots a_{p-1}} = V^{b}\phi_{b a_1\cdots a_{p-1}}$ defines the interior 
contraction of a vector $V$ into a $p$-form $\phi$. Notice that the background data $X_2$, 
$A^{(0)}$ and $\omega_{\mu\nu}$ in (\ref{KSEchi})--(\ref{Dilvarphi}) does not enter equations  (\ref{scalarbil})--(\ref{Omegabil}): they simply drop 
out (one only needs to use the reality properties we specified, together with
the fact that $\omega_{\mu\nu}=\omega_{\nu\mu}$ is symmetric). 

It is straightforward to verify that (\ref{scalarbil})--(\ref{Omegabil}) are invariant under the Weyl transformations
\bea\label{weyl}
 \alpha  &\rightarrow & \Lambda^{-1} \alpha~, \qquad a  \ \rightarrow \  a~, \qquad \ \  K_2 \ \rightarrow \  K_2 - \frac{3}{\sqrt{2}} \dd \log \Lambda~, \nn\\
 S &\rightarrow & \Lambda S~, \qquad \ K_1 \ \rightarrow \  \Lambda K_1~, \qquad b \ \rightarrow \  \Lambda b~,\nn\\
 g &\rightarrow & \Lambda^2 g~, \ \  \qquad J \ \rightarrow \ \Lambda^2 J~, \qquad \Omega \ \rightarrow \  \Lambda^2 \Omega~.
\eea
This symmetry is of course inherited from invariance under the change of radial 
variable $r\rightarrow \Lambda r$ in the bulk.
If $S$ is nowhere zero notice that one might use this symmetry to set $S\equiv 1$. 

Using equations (\ref{scalarbil})--(\ref{Omegabil}) one can show that the conformal Killing vector 
$\xi$ preserves all of the background geometric structure, provided one rescales 
the fields by appropriate powers of $S$ according to their Weyl weights in (\ref{weyl}).
For instance, contracting $\xi$ into the second equation in (\ref{scalarbil}) 
shows that $\mathcal{L}_{\xi}(S\alpha)=0$. On the other hand, taking the 
exterior derivative of the same equation one finds $\mathcal{L}_{\xi}\diff a=0$.  One can hence 
locally choose a gauge in 
which $a$ is invariant under $\mathcal{L}_\xi$, so that the second equation in (\ref{scalarbil}) 
is solved by
\bea\label{agauge}
S\alpha &=& \frac{1}{2\sqrt{2}}i_{\xi}a~.
\eea
In a similar way, one can show 
that also $S^{-1}b$ and  $S^{-2}J$  are invariant under $\mathcal{L}_\xi$, while $S^{-2}\Omega$ is invariant under
$\mathcal{L}_\xi$ in the gauge choice for which (\ref{agauge}) holds. 
Notice that the first equation in (\ref{scalarbil}) implies  that $i_\xi K_2 = -\frac{3}{\sqrt{2}}\mathcal{L}_\xi (\log S)$.

Without loss of generality 
it is  convenient  to henceforth impose  $\mathcal{L}_{\xi}S=0$.\footnote{An exception being 
the $S^1\times S^4$ geometry discussed in section \ref{SecExamples}.}
In terms of the bulk expansion in section \ref{SecHol} this
means choosing the radial coordinate $r$ to be independent of the bulk Killing vector.
This is a natural choice, which
in turn implies that  $\mathcal{L}_{\xi}S=0$ and $SK_1$ is Killing, 
and we shall make this convenient (partial) conformal gauge choice in the following.
We may then introduce a local coordinate $\psi$ so that
\bea
\xi &=& \partial_\psi~.
\eea
The condition $\mathcal{L}_{\xi}S=0$ is then equivalent to $S$ being independent of $\psi$.

\subsection{Geometric structure}\label{SecGeometricStructure}

The Killing vector $\xi$ has norm $S$, and the dual one-form $K_1$ may be written locally
as
\bea
K_1 &=& S(\diff \psi + \rho) \ \equiv \ S\eta~,
\eea
where $i_\xi\rho=0$. Notice that $\eta$ has Weyl weight zero and norm $1/S$.
The local frame $e_1,e_2,e_3,e_4$ provide a basis for $D=\ker \eta$, and $D$ inherits an almost complex structure from $J$. One then defines an endomorphism $\Phi$ of 
the tangent bundle of $M_5$ by
\bea
\Phi\mid_D  &=& I~, \qquad \Phi\mid_{\xi} \ = \ 0~,
\eea
where $I$ is the almost complex structure. One easily verifies that $\Phi^2 = -1 + \xi\otimes \eta$, which is a  defining relation 
of an \emph{almost contact structure}. Moreover, the five-dimensional metric takes the form
\bea
\diff s^2_{M_5} &=& S^2\eta^2 + \diff s^2_4~,
\eea
where $\diff s^2_4$ is Hermitian with respect to $I$. 
Although $\xi$ is Killing, this structure is in general \emph{not} a K-contact structure, which is 
a stronger condition. In particular the latter requires \cite{Boyer:2008era} that $\diff\eta$ is the fundamental $(1,1)$-form 
$J$ associated to the transverse almost complex structure (which in general is not the case here), 
which in turn implies that $\eta$ is a contact form, {\it i.e.} that $\eta\wedge \diff\eta\wedge \diff\eta$ is a volume form
 (which in general is also not the case here).
Notice that 
since $\xi$ is nowhere zero, its orbits define a foliation 
of $M_5$.

Let us now turn to the differential constraints (\ref{scalarbil})--(\ref{Omegabil}). The 
two equations (\ref{scalarbil}) allow us to write
\bea\label{banda}
b &=& \ii S\eta\wedge \left(K_2+\frac{3}{\sqrt{2}}\diff \log S\right) + b_\perp~, \qquad a \ =\ 2\sqrt{2}S\alpha\eta + a_\perp~,
\eea
where $b_\perp$ and $a_\perp$ are \emph{basic} forms for the foliation defined by $\xi$;
 that is, 
they are invariant under, and have zero interior contraction with, $\xi$. Recall that 
in writing the gauge field in the form in (\ref{banda}) we have made a (partial) gauge choice, as in
(\ref{agauge}). This leaves a residual gauge freedom $a_\perp\rightarrow a_\perp + \diff \lambda$, 
where $\lambda$ is a basic ($\xi$-invariant) function. The equation (\ref{K2bil}) is simply 
equivalent to $b$ being invariant under $\xi$. 

The differential constraint (\ref{K1bil}) reduces to
\bea\label{drho}
\diff\rho &=& \frac{\sqrt{2}}{3S}\left(-\ii *_4 b_\perp + 2\ii b_\perp + 4\alpha J\right)~.
\eea
Here $*_4$ is the Hodge dual with respect to the transverse four-dimensional 
metric $\diff s^2_4$, with volume form $e^1\wedge e^2\wedge e^3\wedge e^4$. 
It is then convenient to introduce 
\bea
b_\perp &=& b^+ + b^-~,
\eea
decomposing into the transversely self-dual and anti-self-dual parts. 
Equation (\ref{drho}) is then equivalent to
\bea
b^+ &=& \ii\left(4\alpha J - \frac{3}{\sqrt{2}}S\diff\rho^+\right)~, \qquad b^- \ = \ -\frac{\ii}{\sqrt{2}}S\diff\rho^-~.
\eea

The constraint (\ref{Jbil}) simply identifies
\bea\label{brucelee}
\theta &\equiv & J\hook \diff J \ = \ -\sqrt{2}K_2-\diff \log S~,
\eea
with the \emph{Lee form} $\theta$ of the transverse four-dimensional Hermitian structure. 
That is, every four-dimensional Hermitian structure with fundamental two-form 
$J$ satisfies $\diff J = \theta\wedge J$. Finally, the differential constraint (\ref{Omegabil}) now reads
\bea\label{dOmega}
\diff\Omega &=& (\theta - \ii a_\perp)\wedge \Omega~.
\eea
This implies that the almost complex structure $I$ is \emph{integrable}, thus defining 
 a  \emph{transversely holomorphic foliation}
of $M_5$. We may introduce local coordinates $\psi, z_1,z_2$ adapted 
to the foliation, where the transition functions between the $z_1,z_2$ coordinates 
are holomorphic. 

Notice that we may rewrite (\ref{dOmega}) as
\bea
\diff\Omega &=& -\ii a_{\mathrm{Chern}}\wedge\Omega~,
\eea
where we have defined
\bea
a_{\mathrm{Chern}} &\equiv & a_\perp - I(\theta)~,
\eea
and $I(\theta) \equiv  -i_{\theta^\#}J$, where $\theta^\#$ is the vector field dual to $\theta$. 
To obtain an explicit expression for the Chern connection $a_{\mathrm{Chern}}$, 
we begin by noting
that $\Omega\wedge\bar{\Omega}=2J\wedge J$. 
Using local coordinates $z^\alpha$, $\alpha=1,2,$ for the transverse space
 we may write
\bea
\Omega &=& f\, \diff z^1\wedge \diff z^2~, \qquad J \ =\ \frac{\ii}{2} g^{(4)}_{\alpha\bar\beta}\diff z^\alpha\wedge \diff \bar{z}^{\bar\beta}~,
\eea 
which implies that $|f|=\sqrt{\det g^{(4)}}$. Notice that globally $f$ is a section 
of $L^{-1}$, where $L\cong \Lambda^{2,0}\equiv \mathcal{K}$ is the canonical bundle. 
Writing $f=|f|\ex^{\ii \phi}$ we then have
\bea
\diff\Omega &= & \diff \log f\wedge \Omega \ = \ \ii \, \left(\frac{1}{2} \diff^c \log \det g^{(4)} + \diff\phi\right) \wedge \Omega~,
\eea
where $\diff^c \equiv I\circ \diff$. We thus recognize (up to gauge)
\bea\label{aChern}
a_{\mathrm{Chern}} &=& -\frac{1}{2}\diff^c \log \det g^{(4)}
\eea
as the Chern connection on the canonical bundle.
 
The geometric content of the differential constraints (\ref{scalarbil})--(\ref{Omegabil}) may hence be summarized as follows. 
$M_5$ is equipped with a  transversely Hermitian structure, so that 
the metric takes the form
\bea
\diff s^2_{M_5} &=& S^2 (\diff \psi+\rho)^2 + \diff s^2_4~.
\eea
Here the Killing vector is $\xi=\partial_\psi$, which generates a transversely holomorphic foliation. 
The almost contact form is $\eta=(\diff\psi+\rho)$,
and $\diff s^2_4$ is a transverse Hermitian metric. One is also free to specify 
the functions $\alpha$ and $S$. Given this data, the remaining background fields $a$ and $b$ that enter 
(\ref{scalarbil})--(\ref{Omegabil})
are determined via
\bea\label{aandb}
a &=& 2\sqrt{2}S\alpha\eta +a_{\mathrm{Chern}} + I(\theta)~, \nn\\
b &=& -\frac{\ii}{\sqrt{2}}S\eta\wedge \left(\theta-2\, \diff \log S\right) + 4\ii \alpha J -\frac{\ii}{\sqrt{2}}S(3\diff\rho^+ + \diff\rho^-)~.
\eea
In particular the choice of a transverse Hermitian metric $g^{(4)}$ fixes the 
two-form $J$, and hence the Lee form $\theta$, 
while the Hodge type $(2,0)$-form $\Omega$ and Chern connection $a_{\mathrm{Chern}}$ in (\ref{aChern}) 
are also determined up to gauge. Notice that the terms $S\alpha\eta$ and $I(\theta)$ entering 
the formula for $a$ in (\ref{aandb}) are both global one-forms on $M_5$, implying that 
globally $a$ is a connection on $L=\Lambda^{2,0}$.

We shall furthermore show in section \ref{SecSuff} that \emph{any} choice 
of transversely Hermitian structure on $M_5$ of the above form 
gives a supersymmetric background. In particular the remaining background fields 
$X_2$, $A^{(0)}$, and $\omega_{\mu\nu}$  appearing in the spinor equations 
(\ref{KSEchi})--(\ref{Dilvarphi}) are also determined by the above geometric data. 

\subsection{Examples}\label{SecExamples}

In this section we shall present some explicit examples of the above 
construction. These include all explicit examples appearing in the literature (within 
the Abelian truncation on which we are mostly focusing), 
including examples with six-dimensional gravity duals, plus large families of new solutions. 
 
\subsubsection*{General families}

We begin by noting some special families of backgrounds:
\begin{itemize}
\item Setting $\rho=0$ and $S\equiv 1$ gives a product metric $M_5=\R\times M_4$ or
$M_5 = S^1\times M_4$, 
where $M_4$ is any Hermitian four-manifold. Notice this four-manifold geometry is the same as the rigid supersymmetric 
geometry one finds in four dimensions \cite{Klare:2012gn,  Dumitrescu:2012ha}. 
The first reference here follows a similar holographic approach to the present paper, while 
the second takes a rigid limit of ``new minimal'' supergravity in four dimensions.
\item If $\diff\theta=0$ then the transverse Hermitian metric is 
locally conformally K\"ahler. 
\begin{itemize}
\item If furthermore $\theta=0$ then the transverse four-metric is K\"ahler.
\item If $\theta=0$ and $\diff\rho$ is a positive constant multiple of $J$ then the five-metric is locally conformally Sasakian. 
Supersymmetric gauge theories on Sasaki-Einstein manifolds, 
for which furthermore $S\equiv 1$ and $g$ is a positively curved Einstein metric, 
were defined  in \cite{Qiu:2013pta}, and further studied in \cite{Schmude:2014lfa, Qiu:2013aga,   Qiu:2014oqa, Qiu:2014cha}.
\end{itemize}
\item We may take any circle bundle over a product of Riemann surfaces 
$S^1\hookrightarrow \Sigma_1\times \Sigma_2$. The Hermitian metric 
may be taken to be simply a product of two metrics on the Riemann surfaces, while 
$\rho$ is the connection one-form for the fibration. One can generalize 
this further by allowing $S^1$ orbibundles over a product of 
orbifold Riemann surfaces.
\begin{itemize}
\item If we only fibre over $\Sigma_1$, this leads to direct product $M_3\times \Sigma_2$ 
solutions, where $M_3$ is a Seifert fibred three-manifold. 
Notice this three-manifold geometry is the same as the rigid supersymmetric 
geometry in three dimensions \cite{Closset:2012ru}. Maximally supersymmetric 
Yang-Mills theory has been studied on similar backgrounds 
in \cite{Kawano:2012up, Fukuda:2012jr, Yagi:2013fda, Lee:2013ida, Cordova:2013cea}, 
including the direct products $S^3\times \Sigma_2$ and $M_3\times S^2$. Here the 
spheres are equipped with round metrics and the associated canonical spinors, while the spinors 
on $\Sigma_2$ and $M_3$ are constructed by topologically twisting with 
the $SU(2)_R$ symmetry. 

\end{itemize}
\item Finally, if $\diff\rho$ has Hodge type $(1,1)$ the transversely holomorphic 
foliation admits a complexification \cite{Haefliger}, {\it i.e.} adding a radial direction to $\xi$ 
we naturally have a complex six-manifold $M_6$, with a transversely holomorphic 
foliation. Notice that Sasakian geometry and the direct product $S^1\times M_4$ are special cases. When the orbits of $\xi$ all close, $M_5$ fibres over a Hermitian four-orbifold $M_4$, 
and the associated $U(1)$ orbibundle is the unit circle in a Hermitian holomorphic line 
orbibundle over $M_4$. The corresponding complex $M_6$ is then simply the total space of the associated 
$\C^*$ bundle over $M_4$. 
\end{itemize}

\subsubsection*{Squashed Sasaki-Einstein}

We have already noted that a Sasakian five-manifold is a particular case of a supersymmetric 
background. Recall that Sasakian metrics take the form
\bea\label{Sas}
\diff s^2_5 &=& \eta^2 + \diff s^2_4~,
\eea
where $\eta$ defines a contact structure on $M_5$, with Reeb Killing vector field $\xi$, 
and $\diff s^2_4$ is a transverse K\"ahler metric. Moreover $\diff\eta = \diff\rho = 2J$. 
If the transverse K\"ahler metric $g^{(4)}$ is Einstein, then the metric (\ref{Sas}) is 
said to be a \emph{squashed} Sasaki-Einstein metric.\footnote{In the mathematical literature \cite{Boyer:2008era}
these are called $\eta$-Sasaki-Einstein metrics.} 
For a given choice of transverse K\"ahler-Einstein metric, 
we obtain a two-parameter family of backgrounds, parametrized by 
the constants $c_1,c_2$:
\bea\label{SEsol}
S &\equiv & 1~, \qquad \alpha \ = \ c_1~,
 \qquad K_2 \ = \ -\frac{1}{\sqrt{2}}\theta \ \equiv \ 0~,\nn\\
a & =& c_2 \eta~, \qquad  b \ = \ \ii (4c_1-3\sqrt{2}) J~.
\eea
The K\"ahler-Einstein metric $g^{(4)}$ satisfies the Einstein equation $\mathrm{Ric}^{(4)} = 2(2\sqrt{2}c_1-c_2)g^{(4)}$. 
Notice that we have presented the solution (\ref{SEsol}) in a different gauge choice to (\ref{agauge}). 
We may impose the latter gauge choice by simply transforming 
$a\rightarrow a + (2\sqrt{2}c_1-c_2)\diff\psi$, although the form of $a$ in (\ref{SEsol}) makes it clear that 
we may take $a$ to be a global one-form on $M_5$ for this particular class of solutions.

When $g_4$ is taken to be the standard metric on $\mathbb{CP}^2$, the above geometry is a 
squashed five-sphere. This corresponds to the conformal boundary of the 1/4 BPS bulk Romans supergravity 
solutions constructed in \cite{Alday:2014bta}. 

\subsubsection*{Black hole boundary}

In this section we consider the conformal boundary of the 1/2 BPS topological black hole 
solutions constructed in \cite{Alday:2014fsa}. 
We begin with the following product metric on $S^1\times \mathbb{H}^4$, where 
$\mathbb{H}^4$ is hyperbolic four-space:
\bea
\dd s^2_5 &=&\diff \tau^2+ \frac{1}{q^2+1}\diff q^2 + q^2(\diff\vartheta^2+\sin^2\vartheta \diff\varphi_1^2 
+ \cos^2\vartheta \diff\varphi_2^2)~.
\eea
Here $\tau$ is a periodic coordinate on $S^1$, $q$ is a radial coordinate with $q\in[0,\infty)$, 
$\vartheta\in [0,\frac{\pi}{2}]$ while $\varphi_1, \varphi_2$ have period $2\pi$. 
The metric in brackets is simply the round metric on a unit radius $S^3$. 
For this solution $b$ vanishes identically, while $a$ is gauge-equivalent to zero. 
The Killing spinors for this background \cite{Alday:2014fsa} in general depend on four integration constants 
(being 1/2 BPS), but for simplicity here we present only the ``toric'' solution discussed in \cite{Alday:2014fsa}.
The remaining fields are then
\bea
S &=& \sqrt{q^2+1}~, \qquad \alpha \ = \ -\frac{3}{2\sqrt{2}\sqrt{q^2+1}}~,\nn\\
K_2 &=& -\frac{3}{\sqrt{2}}\frac{q}{q^2+1}\diff q \ =\ -\frac{3}{2\sqrt{2}}\diff \log (q^2+1)~,
\eea
while in a gauge\footnote{This is different to the gauge choice (\ref{agauge}), where instead $a=-3\diff\tau$ for this solution.} in which $a=0$
the $U(2)$ structure is given by
\bea
K_1 &=& \frac{1}{\sqrt{q^2+1}}\left[\diff \tau +q^2(\cos^2\vartheta\diff\varphi_2 - \sin^2\vartheta\diff\varphi_1)\right]~,\nn\\
 J  &=& \ \frac{q^2 }{2} \, \sin 2 \vartheta \, \diff\vartheta\wedge (\diff\varphi_1+\diff\varphi_2) + \frac{q }{\left(q^2+1\right)} \diff q\wedge \bigg[  \diff\tau + \sin ^2\vartheta \diff\varphi_1-\cos ^2\vartheta  \diff\varphi_2\bigg] \, ,\nn\\
\Omega &=&  -\frac{q \, \ex^{\ii (\varphi_1-\tau -\varphi_2)}}{2 \sqrt{q^2+1}} \bigg[\sin 2 \vartheta\, 
(q\, \diff\tau - \ii \diff q)\wedge (\diff\varphi_1+\diff\varphi_2) + q \, \sin 2 \vartheta\,  \diff\varphi_1\wedge \diff\varphi_2 \nn\\
&&+ 2\ii\, q\, \diff\vartheta\wedge ( \diff\tau + \sin ^2\vartheta \diff\varphi_1-\cos ^2\vartheta  \diff\varphi_2) - 2\, \diff q \wedge \diff\vartheta \bigg]~.
\eea
The supersymmetric Killing vector is 
\bea
\xi &=& g(SK_1,\, \cdot\, ) \ = \ \partial_\tau + \partial_{\varphi_2}-\partial_{\varphi_1}~.
\eea
Furthermore, notice that rescaling $J$ by $1/(q^2+1)$ leads to a closed two-form, 
hence showing that the Hermitian metric transverse to $\xi$ is conformal to a K\"ahler metric.
Moreover, one can also check that the almost contact form $\eta=K_1/S$ is a contact 
form in this case, {\it i.e.} that $\eta\wedge\diff\eta\wedge \diff\eta$ is a volume form. 

\subsubsection*{Conformally flat $S^1\times S^4$}

In this section we consider the conformally flat metric on $S^1\times S^4$, which 
we may write as
\bea\label{S1S4}
\diff s^2_5 &=& \diff\tau^2 + \diff s^2_{S^4}~,
\eea
where
\bea\label{S4}
\diff s^2_{S^4} &=& \diff\beta^2 + \sin^2\beta(\diff\vartheta^2+\sin^2\vartheta \diff\varphi_1^2 
+ \cos^2\vartheta \diff\varphi_2^2)~.
\eea
Here $\tau$ is a  periodic coordinate on $S^1$, while the metric in brackets in (\ref{S4}) is simply the round metric on a unit radius $S^3$, as in the previous black hole boundary example. The polar coordinate $\beta\in[0,\pi]$. 
The metric (\ref{S1S4}) of course arises as the conformal boundary of Euclidean AdS in global coordinates, 
and as such the background fields $a=0=b$. There are many Killing spinors in this case, 
and here we simply choose one so as to present simple expressions for the remaining 
background data. We find
\bea
S &=&  \ex^{-\tau}~, \qquad \alpha \ = \ 0~, \qquad K_2 \ = \ \frac{3}{\sqrt{2}}\diff\tau~.
\eea
The $U(2)$ structure is given by
\bea
K_1 &=& \sin\beta\, \diff\beta - \cos\beta\, \diff\tau~,\nn\\
J &=& \sin^2 \beta  \sin (\varphi_1 +\varphi_2 ) \bigg\{ \cot (\varphi_1 +\varphi_2 ) \left( \diff \vartheta \wedge \diff \tau- \cot \beta\,  \diff \beta\wedge \diff \vartheta\right)- \sin ^2\vartheta \, \diff \vartheta \wedge \diff \varphi_1\nn\\
&& -\cos ^2\vartheta \, \diff \vartheta \wedge \diff \varphi_2 +
\sin \vartheta \cos \vartheta\Big[ (\cot \beta \, \diff \beta+ \diff \tau ) \wedge (\diff \varphi_1 - \diff \varphi_2)\nn\\
&& -\cot (\varphi_1+\varphi_2 ) \, \diff \varphi_1\wedge \diff \varphi_2 \Big]\bigg\} \, ,\nn\\
\Omega &
=  & \ii \sin^2 \beta\sin (\varphi_1 +\varphi_2 ) \Big[ \cot \beta  \, \diff \beta\wedge \diff \vartheta - \, \diff \vartheta \wedge \diff \tau + \sin \vartheta \cos \vartheta \,  \diff \varphi_1\wedge \diff \varphi_2 \Big]   \nn \\
&&+\sin^2 \beta \sin \vartheta \Big[\sin \vartheta+\ii \cos \vartheta \cos (\varphi_1 +\varphi_2 )\Big] \Big( \cot \beta  \, \diff \beta\wedge \diff \varphi_1 - \cot \vartheta \,  \diff \vartheta \wedge \diff \varphi_2\nn\\
&&+ \diff \tau \wedge \diff \varphi_1\Big) 
+\sin^2 \beta \cos \vartheta \Big[\cos \vartheta -\ii \sin \vartheta \cos (\varphi_1 +\varphi_2 )\Big] \Big( \cot \beta  \,  \diff \beta\wedge \diff \varphi_2\nn\\
&&+ \tan \vartheta \, \diff \vartheta \wedge \diff \varphi_1+  \, \diff \tau \wedge \diff \varphi_2 \Big) \, .
\eea
Notice that in this example we obtain a  \emph{conformal} Killing vector 
from the Killing spinor bilinear, but not a Killing vector.
 As described at the end of section \ref{SecDiffConstraints}, we may 
always make a Weyl transformation of the background to obtain a Killing vector. In the case 
at hand this corresponds to the Weyl factor $\Lambda=\ex^{\tau}$, and the corresponding 
Weyl-transformed metric is then (locally) flat, with the Weyl-transformed $J$ and $\Omega$ both closed and hence 
defining a transverse hyperK\"ahler structure. Nevertheless, the fact 
that the metric (\ref{S1S4}) leads to a conformal Killing vector explains why 
this background is missing from the rigid supersymmetric geometry in \cite{Pan:2013uoa, Imamura:2014ima}:
in the latter references the corresponding bilinear is necessarily a Killing vector. 
This also suggests that the  conjecture made in \cite{Imamura:2014ima} is likely to be correct: that is, 
to obtain the $S^1\times S^4$ background from a rigid limit of supergravity, one should begin with 
\emph{conformal} supergravity in five dimensions, rather than Poincar\'e supergravity.\footnote{See \cite{Pini:2015xha} for a related discussion on this point.} 

\subsubsection*{Squashed $S^5$}

We consider the squashed five-sphere metric 
\bea
\diff s^2_5 &=& \frac{1}{s^2}(\diff\tau+C)^2+ 
\diff\sigma^2 + \frac{1}{4}\sin^2\sigma(\diff\vartheta^2+\sin^2\vartheta\dd\varphi^2)\nn\\
&&+\frac{1}{4}\cos^2\sigma\sin^2\sigma (\dd\beta+\cos\vartheta\dd\varphi)^2~,
\eea
where $s\in (0,1]$ is the squashing parameter and
\bea\label{C}
C &\equiv & - \frac{1}{2}\sin^2\sigma(\dd\beta+\cos\vartheta\dd\varphi)~.
\eea
The coordinates $\sigma, \beta, \vartheta, \varphi$ are 
coordinates on the base $\mathbb{CP}^2$, with $\beta$ having period $4\pi$, 
$\varphi$ having period $2\pi$,  while $\sigma\in [0,\tfrac{\pi}{2}]$, $\vartheta\in[0,\pi]$, and $\frac{1}{2}\diff C$ is the K\"ahler two-form on $\mathbb{CP}^2$. For the ``toric'' family discussed in \cite{Alday:2014rxa, Alday:2014bta} we find
\bea
 S & = &   \frac{\cos^2 \sigma}{b_2}+ \frac{\sin^2 \sigma}{b_1}~,
 \eea
 where 
 \bea
  b_1  \ = \  1+ \sqrt{1-s^2} \, , \qquad b_2 \ = \  1- \sqrt{1-s^2} \, .
 \eea
The other background fields are, in an appropriate gauge ({\it i.e.} not that in (\ref{agauge})),
\bea
 \alpha & = &  \frac{b_1 (b_1+b_2) (b_1 - 7b_2 +(b_1-b_2) \cos 2\sigma)}{ 4 \sqrt{2} (b_1 \cos^2 \sigma + b_2 \sin^2 \sigma)}~,\nn\\
  a & =  & \frac{b_1-b_2}{2 b_2} \left( \diff \tau+C \right) ~\nn,\\
   b&=&  -  \frac{\ii (b_1 - b_2)}{2 \sqrt{2} b_1 b_2 (b_1 + b_2)}\diff C~, \nn\\
    K_2 & =  & \frac{\sqrt{2} \ (b_1-b_2) \ \sin 2 \sigma}{b_1 \cos^2 \sigma + b_2 \sin^2 \sigma} \ \diff \sigma \ = \ -\sqrt{2}\, \diff \log \, (b_1\cos^2\sigma + b_2\sin^2\sigma)~.
 \eea
 The $U(2)$ structure is
 \bea
  K_1 & =  & \ \frac{1}{4 b_1 b_2(b_1+b_2) \big(b_1\cos^2\sigma+b_2\sin^2\sigma\big)} \bigg[(b_1+b_2)(b_1-b_2+(b_1+b_2)\cos 2\sigma)\diff \tau\nn\\
&& -\frac{1}{2}\sin^2 \sigma \, \Big((b_1-b_2)^2 \cos 2 \sigma+ b_1^2-4 b_1 b_2-b_2^2\Big) (\diff \beta+\cos \vartheta  \diff \varphi)\bigg]
 \, ,\nn \\
 J  & =  & \  \frac{ \sin \sigma}{8 \, b_1 b_2 \, (b_1+b_2)^2  \, \left(b_1 \cos ^2\sigma+b_2 \sin ^2\sigma\right)}  \, \bigg[4 \cos \sigma \, \Big( \, 2 (b_1+b_2) \, \diff\sigma\wedge \diff\tau \nn\\
 && - b_1 \diff\sigma\wedge(\diff\beta+\cos\vartheta\diff\varphi)\Big)+ 2\sin \vartheta \sin \sigma \, (b_1 \cos ^2\sigma+b_2 \sin ^2\sigma)  \diff\vartheta\wedge \diff\varphi \bigg]\, ,\nn\\
 \Omega & = & \ \frac{\sin \sigma \, \ex^{\ii (\tau -\beta )} }{8  b_1 b_2 (b_1+b_2)^2 \left(b_1 \cos ^2\sigma+b_2 \sin ^2\sigma\right)} \, \bigg[-\sin 2 \sigma \, \Big(\ii \sin \vartheta \, ( \, b_1 \diff\varphi\wedge \diff\beta \nn\\
 && + 2 (b_1+b_2) \, \diff\tau\wedge \diff\varphi)- 2 (b_1+b_2) \, \diff\vartheta\wedge \diff\tau +b_1\, \diff\vartheta\wedge (\diff\beta+\cos\vartheta\diff\varphi)\Big)\nn\\
 &&  -4\, (b_1 \cos ^2\sigma+b_2 \sin ^2\sigma) \, (\sin \vartheta \ \diff\sigma\wedge \diff\varphi + \ii \, \diff\vartheta\wedge \diff\sigma) \bigg] \, .
 \eea
 The supersymmetric Killing vector is
 \bea
 \xi &=& b_1\partial_\tau + 2(b_1+b_2)\partial_\beta~.
 \eea
One also computes
\bea\label{etanotcontact}
\eta\wedge\diff\eta\wedge \diff\eta &=& \frac{ b_1^3 b_2^3 \  (b_1+b_2)^2}{2\left( b_1\cos^2\sigma+b_2\sin^2\sigma\right)^5} \left((b_1-b_2)^2 \cos 2 \sigma +b_1^2-4 b_1 b_2-b_2^2\right) \nn \\
 &&\times \left(\left(b_1^2-b_2^2\right) \cos 2 \sigma+b_1^2-6 b_1 b_2+b_2^2\right) \vol_5 \, ,
\eea
where $\vol_5$ denotes the Riemannian volume form and $\eta=K_1/S$ is the almost contact form. The right hand side of (\ref{etanotcontact}) can have non-trivial zeros, and we thus see that 
in general $\eta$ does not define a contact structure. 
 These backgrounds arise as the conformal boundary of the 3/4 BPS solutions of Romans supergravity 
 constructed in \cite{Alday:2014rxa, Alday:2014bta}. 

\subsection{From geometry to supersymmetry}\label{SecSuff}

In this section we will show 
that \emph{any} choice 
of transversely Hermitian  structure on $M_5$
defines a supersymmetric background. The background 
$U(1)_R$ gauge field $a$ and the $b$-field are given in terms of the 
geometry by (\ref{aandb}). It then remains to show that the geometry 
also determines the fields $X_2$, $A^{(0)}$ and $\omega_{\mu\nu}$, 
in such a way that the original spinor equations (\ref{KSEchi})--(\ref{Dilvarphi}) 
are satisfied.

We first examine the Killing spinor equation (\ref{KSEchi}) for $\chi$.  In order to proceed it is convenient to choose a set of projection conditions 
(see for example \cite{Gauntlett:2003cy})
\bea\label{proj}
\gamma_{12}\chi \ = \ \gamma_{34}\chi \ = \ \ii \chi~, \qquad \gamma_5\chi &=& \chi~.
\eea
These allow one to substitute for the fields $b$ and $K_2$ in terms of the geometry, 
via (\ref{aandb}) and (\ref{brucelee}), into the right hand side of equation (\ref{KSEchi}).  
In doing this calculation it is also convenient to 
write $\Omega=J_2+\ii J_1$, $J=J_3$ so that 
\bea
J_1 &=& e_{14}+e_{23}~, \qquad J_2 \ = \ e_{13}-e_{24}~, \qquad J_3 \ = \ e_{12}+e_{34}~.
\eea
Notice that $J_i$, $i=1,2,3$ span the transverse self-dual two-forms, and 
hence may  be used as a basis thereof. One can furthermore make use of various identities that 
easily follow from (\ref{proj}), such as $\ii \gamma_m\chi = J_{mn}\gamma^n\chi$, where $m,n=1,\ldots,4$, 
and $(\beta^-)_{mn}\gamma^{mn}\chi=0$ for any transverse anti-self-dual two-form $\beta^-$. 

In this way
it is straightforward to show that the $\mu=5$ (the $\psi$ direction) component of (\ref{KSEchi}) 
simply imposes $\partial_\psi \chi=0$.\footnote{Without loss of generality we  take the four-dimensional 
frame $e^1,\ldots,e^4$ to be independent of the Killing vector $\xi=\partial_\psi$.} Thus 
$\chi$ is independent of $\psi$.  Taking instead $\mu=m$, $m=1,2,3,4$, one finds (\ref{KSEchi}) 
is equivalent to
\bea\label{4dKSE}
\nabla^{(4)}_m\chi &=& \frac{1}{4}\theta^n\gamma_{mn}\chi - \frac{\ii}{2}(a_\perp)_m \chi + \frac{1}{2}(\partial_m\log S) \chi~, 
\eea
where $\nabla^{(4)}$ denotes the Levi-Civita spin connection for the 
transverse four-dimensional metric. Recall that the latter metric is Hermitian.
It is then more natural to express  equation (\ref{4dKSE}) in terms of an appropriate Hermitian connection, 
which preserves both the metric and the two-form $J$. The Chern connection is such a connection, 
defined by
\bea
\nabla^{\mathrm{Chern}}_m\chi &=& \partial_m\chi + \frac{1}{4}(\omega^{\mathrm{Chern}}_{m})_{pq}\gamma^{pq}\chi~, \nn\\
\mbox{where} \qquad (\omega^{\mathrm{Chern}}_m)_{pq} & \equiv &  (\omega^{(4)}_m)_{pq}+\frac{1}{2}J_{m}^{\ n}(\diff J)_{npq}~.
\eea
This coincides with the Levi-Civita connection if and only if $\diff J=0$ (equivalently $\theta=0$), so that the 
metric is K\"ahler.

Next, let us notice that under the Weyl transformation 
(\ref{weyl}) we have $\chi\rightarrow \Lambda^{1/2}\chi$,  so that it is also natural to introduce 
\bea
\tilde{\chi} & \equiv & S^{-1/2}\chi~,
\eea 
so that $\tilde{\chi}$ is Weyl invariant. In this notation (\ref{4dKSE}) becomes
\bea\label{Cherny}
\nabla^{\mathrm{Chern}}_m \tilde\chi + \frac{\ii}{2}a_{\mathrm{Chern}} \tilde\chi  &=& 0~,
\eea
where recall that $a_{\mathrm{Chern}}=a_\perp - I(\theta)$ is the Chern connection for the 
canonical bundle $\mathcal{K}\equiv \Lambda^{2,0}$, given explicitly by (\ref{aChern}). It is then a standard fact, and is straightforward to show, 
that any Hermitian space admits a canonical solution $\tilde\chi$ to (\ref{Cherny}). 
Specifically, any Hermitian space admits a canonical spin$^c$ structure, 
with twisted spin bundles $\mathrm{Spin}^c=\mathrm{Spin}\otimes \mathcal{K}^{-1/2}$. 
In four dimensions this is  isomorphic to
\bea\label{spinc}
\mathrm{Spin}^c &\cong & \left(\Lambda^{0,0}\oplus \Lambda^{0,2}\right)\oplus \Lambda^{0,1}~,
\eea
where $\Lambda^{p,q}$ denotes the bundle of forms of Hodge type $(p,q)$. In the case at hand, 
these are defined transversely to the foliation generated by the Killing vector $\xi$. Under (\ref{spinc})
the Killing spinor $\tilde\chi = S^{-1/2}\chi$ is a section of the trivial line bundle 
$\Lambda^{0,0}$. Moreover, the Chern connection restricted to this summand is flat, 
with the induced connection $-\frac{1}{2}a_{\mathrm{Chern}}$ on the twist factor $\mathcal{K}^{-1/2}$ 
effectively cancelling that coming from the spin bundle. Concretely, in terms of local 
complex coordinates $z^\alpha$, $\alpha=1,2$, we have $(\omega^{\mathrm{Chern}})_\alpha^{\ \beta} = 
(\partial g^{(4)})_{\alpha\bar\gamma}(g^{(4)})^{\bar\gamma\beta}$, 
and using the projection conditions (\ref{proj}) one can show this precisely cancels 
the contribution from (\ref{aChern}). 
The spin$^c$ spinor $\tilde\chi$ is simply a constant length section of this flat line bundle. 
Put simply, the rescaled Killing spinor $\tilde\chi = S^{-1/2}\chi$ is constant. 

Next we turn to the dilatino equation (\ref{Dilchi}). 
Substituting for $\varphi$ in terms of $\chi$, 
using (\ref{Dilvarphi}), after a somewhat lengthy computation one finds the dilatino 
equation holds provided
\bea\label{A0}
A^{(0)} &= & -\frac{9}{4}*\left(\diff *b - \frac{\ii \sqrt{2}}{3}b\wedge b\right)~,
\eea
and
\bea
X_2 \ = \ -4\alpha^2-\frac{1}{4}\langle K_2, K_2\rangle -\frac{\ii}{6\sqrt{2}}S\langle \eta, A^{(0)}\rangle 
-\frac{3}{16}\langle \diff a_\perp, J \rangle - \frac{3}{4\sqrt{2}}\langle K_2,\diff\log S\rangle~.
\eea
Here we have introduced the notation $\phi_1\wedge *\phi_2 = \frac{1}{p!}\langle \phi_1,\phi_2\rangle\, \vol_5$ 
for the inner product between two $p$-forms $\phi_1$, $\phi_2$. 
Notice that the expression (\ref{A0}) for the imaginary one-form $A^{(0)}$ 
coincides with that in \cite{Alday:2014bta}, which was derived 
by solving the bulk equations of motion near
the conformal boundary, in terms of the boundary data. 
Notice that under the Weyl scaling (\ref{weyl}) we have
\bea\label{weylA}
A^{(0)} & \rightarrow & \frac{1}{\Lambda}\left(A^{(0)} + \frac{9}{2}i_{\diff \log \Lambda^\#} b \right)~, \qquad 
X_2 \ \rightarrow \frac{1}{\Lambda^2}X_2~.
\eea
The fact that $X_2$ has Weyl weight $-2$ is clearly consistent with the bulk expansion 
(\ref{expand}), but the ``anomalous'' transformation of $A^{(0)}$ in (\ref{weylA}) 
 naively appears to contradict (\ref{expand}), for which $A^{(0)}$ has Weyl 
weight $-1$. However, this is where the comment above equation (\ref{expand}) 
is relevant: the reparametrization $r\rightarrow \Lambda r$ does \emph{not} 
preserve the \emph{subleading} terms in the metric (\ref{diagonalmetric}). 
It is therefore not a strict symmetry of the system we have defined.
However, the \emph{leading order} terms in the expansions (\ref{diagonalmetric}), (\ref{expand}) 
\emph{are} invariant. This explains why the differential constraints  (\ref{scalarbil})--(\ref{Omegabil}) 
have the Weyl symmetry (\ref{weyl}), while the higher order term $A^{(0)}$ arising  in the expansion 
of the $B$-field does not. One could restore the full Weyl symmetry by adding a 
cross term $9\frac{\diff r}{r}\Cweyl_\mu \diff x^\mu$ into the metric (\ref{diagonalmetric}), 
so that 
\bea\label{WeylC}
\Cweyl & \rightarrow & \Cweyl-\diff\log \Lambda~,
\eea
 under $r\rightarrow \Lambda r$ preserves the form of the metric.
Then $\Cweyl$ is a new background field on $M_5$, and one finds
\bea\label{weylA0}
A^{(0)} &= & -\frac{9}{4}*\left[(\diff+2\Cweyl\wedge) *b - \frac{\ii \sqrt{2}}{3}b\wedge b \right]~.
\eea
This now has Weyl weight $-1$, as expected, and the anomalous variation in (\ref{weylA}) 
arises simply because we have made the gauge choice $\Cweyl=0$ in our original expansion.
In general notice that a field of Weyl weight $w$ will couple to a Weyl covariant derivative 
$\mathrm{D}_\mu\equiv \partial_\mu + w \, \Cweyl_\mu$, and $w=2$ for $*b$. 

It remains to show that the background geometry implies
the $\varphi$ Killing spinor equation (\ref{KSEvarphi}). At this point notice that 
everything is fixed uniquely in terms of the free functions $\alpha$ and $S$, and 
the transversely Hermitian structure on $M_5$, apart from the higher order 
spin connection term $\omega_{\mu\nu}$ which appears in (\ref{KSEvarphi}). 
After a lengthy computation, in our orthonormal frame one finds the expression
\bea
\omega_{55} &=& -6\sqrt{2}\alpha^2 - \frac{1}{3\sqrt{2}}\langle K_2, K_2\rangle - \sqrt{2}X_2 -\frac{1}{2\sqrt{2}}
\langle \diff a_\perp, J\rangle - \langle K_2,\diff \log S\rangle~,\nn\\
\omega_{5m} &=& \left[-\frac{\ii}{3\sqrt{2}}i_{K_2^\#}b_\perp + i_{\diff\log S^\#}\left(2\alpha J +
\frac{1}{\sqrt{2}}S \diff\rho^-\right)\right]_m \ = \ \omega_{m5}~,\nn\\
\omega_{mn} &=& \frac{\sqrt{2}}{3}(K_2)_m(K_2)_n - {\nabla^{(4)}}_{(m} (K_2)_{n)} - \big(\frac{4}{3}S\alpha \,
\diff\rho^- +\frac{1}{\sqrt{2}}\diff a_\perp^-\big)_{mp}J^{p}_{\ n}\nn\\
&&+\bigg(2\sqrt{2}\alpha^2 + \frac{\sqrt{2}}{3}X_2 - \frac{1}{3\sqrt{2}}\langle K_2,K_2\rangle+ \frac{1}{4\sqrt{2}}\langle \diff \alpha_\perp , J \rangle\bigg)\delta_{mn} ~.\label{omegamunu}
\eea
This is manifestly real and symmetric, apart from the last term in the penultimate line. 
However, it is straightforward to show that $(\beta^-)_{mp}J^{p}_{\ n}$ 
is symmetric for any transverse anti-self-dual two-form $\beta^-$. 
Thus (\ref{KSEvarphi}) is satisfied provided $\omega_{\mu\nu}$ 
is given by (\ref{omegamunu}). We conclude this subsection by noting the 
following formula 
\bea
\omega_\mu^{\ \mu} \ =\  2\sqrt{2}\alpha^2+\frac{\sqrt{2}}{3}X_2 - \langle K_2, \frac{1}{\sqrt{2}}K_2+\diff \log S\rangle + \frac{1}{2\sqrt{2}}
\langle \diff a_\perp, J\rangle  - \nabla^{(4)}_mK_2^m~.
\eea
This trace will appear in the supersymmetric Lagrangians constructed in section \ref{SecLagrangians}.

\subsection{Summary}\label{SecSummary}

A supersymmetric asymptotically locally AdS solution to six-dimensional Romans supergravity 
leads to the coupled spinor equations  (\ref{KSEchi})--(\ref{Dilvarphi}) on the conformal boundary $M_5$. 
These are a rather complicated looking set of equations for the spin$^c$ spinors $\chi$, $\varphi$, depending on the large number of background 
fields $g, X_2, a, A^{(0)}, b$ and $\omega_{\mu\nu}$ on $M_5$, with 
$\varphi$ and $\chi$ related to each other by the further background fields 
$\alpha$ and $K_2$ via (\ref{Dilvarphi}). 
However, we have shown these equations are completely equivalent to a very simple 
geometric structure:

\begin{enumerate}[(i)]
\item The five-manifold $M_5$ is equipped with a transversely holomorphic foliation, 
with the one-dimensional leaves generated by the (conformal) Killing vector field 
$\xi=\partial_\psi$. This structure is a natural odd-dimensional cousin of a complex manifold, 
and means we may cover $M_5$ locally with coordinates $\psi, z_1,z_2$, where 
the transition functions between the $z_1,z_2$ coordinates are holomorphic (more formally 
we have an open cover $\{U_i\}$ and submersions $f_i:U_i\rightarrow \C^2$ 
with one-dimensional fibres, such that on overlaps $U_i\cap U_j$ we have $f_j=g_{ji}\circ f_i$ 
where $g_{ji}$ are biholomorphisms of open sets in $\C^2$).
\item This foliation is compatible with an almost contact form $\eta=\diff \psi + \rho$. 
Choose a particular $\rho=\rho_0$, which notice is defined only locally in the foliation patches, gluing 
together to give the global $\eta$.
Then for fixed foliation any other choice of $\rho$ is related to this by $\rho=\rho_0+\nu$, 
where $\nu$ is a \emph{global basic one-form}. That is, 
$\nu$ is  a global one-form on $M_5$ satisfying $\mathcal{L}_\xi\nu=0=i_\xi\nu$. 
\item One can choose an \emph{arbitrary} transverse Hermitian metric $\diff s^2_4$, 
invariant under $\xi$ and compatible with the foliation.
\item Finally, one is free to choose the $\xi$-invariant real functions $\alpha$ and $S$ (with $S$ nowhere zero). 
\end{enumerate}
An interesting special case is when all the leaves of the foliation are closed, so that $\xi$ generates 
a $U(1)$ action on $M_5$ and $\psi$ is a periodic coordinate. In this case $M_5$ fibres over a complex Hermitian orbifold $M_4=M_5/U(1)$, 
where $\eta$ is a global angular form for the $U(1)$ orbibundle. Different choices of $\nu$ in (ii) above 
are then simply different connections on this bundle, with (iii) giving different Hermitian metrics on $M_4$.

We have shown that any choice of the data (i)--(iv) determines a supersymmetric background, solving 
the spinor equations (\ref{KSEchi})--(\ref{Dilvarphi}), and conversely any such solution 
determines a choice of the above geometric data. Furthermore, solving  (\ref{KSEchi})--(\ref{Dilvarphi}) 
is equivalent to finding a supersymmetric asymptotically locally 
AdS solution to Romans supergravity, to the first few orders in an expansion around the 
conformal boundary $M_5$. Of course whether or not this extends to a complete
non-singular supergravity solution, as some of the explicit examples in section \ref{SecExamples} do,
 is another matter.

%%%%%%%%%%%%%%%%%%%%%%%%%%%%%%%%%%%%%%%%%%%

\section{Supersymmetric gauge theories}\label{SecLagrangians}

In this section we construct $\mathcal{N}=1$ supersymmetric gauge theories formed of vector and hypermultiplets on the background geometry described in section \ref{SecGeometry}. 

We adopt the same notation as \cite{Hosomichi:2012ek}, in particular using $\xi$ and $\eta$ to denote  five-dimensional 
Killing spinors. The $\gamma_\mu$ are $4 \times 4$ Hermitian matrices which form a basis of Cliff$(5,0)$ in an orthonormal frame. A complete set of $4 \times 4$ matrices is given by $(1_4,\gamma_\mu,\gamma_{\mu\nu})$ and we choose $\gamma_{\mu\nu\rho\sigma\tau} = - \epsilon_{\mu\nu\rho\sigma\tau}$ with $\epsilon_{12345} =+1$. The five-dimensional charge conjugation matrix, $\mathcal{C}=(\mathcal{C}_{\alpha\beta})$, is unitary and anti-symmetric in the spinor indices $\alpha,\beta=1,2,3,4$ of $ \mathrm{Spin}(5)\cong Sp(2)$. The matrices $\mathcal{C} \gamma_\mu$ are anti-symmetric in spinor indices whereas $\mathcal{C} \gamma_{\mu\nu}$ are symmetric. Spinor bilinears are denoted $( \eta \gamma^{\mu_1 \cdots \mu_n} \xi ) \ = \ \eta^\alpha ( \mathcal{C} \gamma^{\mu_1 \cdots \mu_n} )_{\alpha\beta} \xi^\beta $. Finally, the Fierz identity for Grassmann odd spinors in five dimensions is
\begin{align}
\gamma^A \eta^\alpha ( \xi \gamma^B \lambda ) \ = \ - \frac{1}{4} ( \eta \xi ) \gamma^A \gamma^B \lambda^\alpha - \frac{1}{4} ( \eta \gamma_\mu \xi ) \gamma^A \gamma^\mu \gamma^B \lambda^\alpha + \frac{1}{8} ( \eta \gamma_{\mu\nu} \xi ) \gamma^A \gamma^{\mu\nu} \gamma^B \lambda^\alpha \, ,
\end{align}
where $\gamma^{A}, \gamma^B$ denote arbitrary elements of Cliff$(5,0)$.

\subsection{Supersymmetry algebra}

An off-shell $\mathcal{N}=1$ vector multiplet in five dimensions consists of a gauge field $\mathcal{A}_\mu$, a real scalar $\sigma$, a gaugino $\lambda_I$, and a triplet of auxiliary scalars $D_{IJ}$, all transforming in the adjoint representation of the gauge group $G$. 
Here $I,J=1,2$ are $SU(2)_R$ symmetry indices. The gaugino is a symplectic-Majorana spinor which satisfies $( \lambda_I^\alpha )^* = \varepsilon^{IJ} \mathcal{C}_{\alpha\beta} \lambda^\beta_J$ whilst the auxiliary scalars satisfy $(D_{IJ})^\dagger = \varepsilon^{IK} \varepsilon^{JL} D_{KL}$, where recall that $\varepsilon^{IJ}$ is the 
Levi-Civita symbol.

We introduce the following covariant derivatives:
\bea
\mathcal{F}_{\mu\nu} &=& \partial_\mu \mathcal{A}_\nu - \partial_\nu \mathcal{A}_\mu - \ii [ \mathcal{A}_\mu , \mathcal{A}_\nu ]~, \nn\\
D_\mu \sigma &=& \partial_\mu \sigma - \ii [ \mathcal{A}_\mu , \sigma ]~, \nn\\
D_\mu \lambda_I &=& \nabla_\mu \lambda_I - \ii [ \mathcal{A}_\mu , \lambda_I ] ~, \nn\\
D_\mu D_{IJ} &=& \partial_\mu D_{IJ} - \ii [ \mathcal{A}_\mu , D_{IJ} ] \, ,
\eea
where $\nabla$ is the Levi-Civita spin connection. 
In general we may consider turning on an $SU(2)_R$ background gauge field $a_\mu^i$, $i=1,2,3$, 
or equivalently we may introduce
\bea
V_{\mu IJ} &\equiv & -\frac{\ii}{2}a_\mu^i(\sigma_i)_{IJ}~,
\eea
where $\sigma_i$, $i=1,2,3$, denote the Pauli matrices. In section \ref{SecHol} recall that for simplicity
we restricted to an Abelian background gauge field, with $a^1=a^2=0$, $a^3=a$, 
but in this section we will relax this assumption. There is also a background two-form $b$-field and we choose to introduce the gauge field $\Cweyl_\mu$ associated with restoring Weyl invariance -- see the earlier discussion around equation (\ref{WeylC}). With this background gauge field active we modify the covariant derivatives to 
\bea
\mathrm{D}_\mu \sigma &=& D_\mu \sigma - \Cweyl_\mu \sigma ~, \nn\\
\mathrm{D}_\mu \lambda_I &=& D_\mu \lambda_I - \frac{3}{2} \Cweyl_\mu \lambda_I - V_{\mu I}{}^J \lambda_J ~, \nn\\
\mathrm{D}_\mu D_{IJ} &=& D_\mu D_{IJ} - 2 \Cweyl_\mu D_{IJ} - 2 V_{\mu (I}{}^K D_{J)K} \, ,
\eea
so that they are covariant with respect to both Weyl and R-symmetry transformations. These correspond to 
Weyl weights $w=(-1,0,-\frac{3}{2},-2)$ for the gauge multiplet $(\sigma,\mathcal{A}_\mu,\lambda_I,D_{IJ})$. 

Given this background data
we consider the following (conformal) supersymmetry variations:
\bea
\delta_\xi \sigma &=& \phantom{-} \ii \varepsilon^{IJ} \xi_I \lambda_J~,\nn\\
\delta_\xi \mathcal{A}_\mu &=& \phantom{-} \ii \varepsilon^{IJ} \xi_I \gamma_\mu \lambda_J~, \nn\\
\delta_\xi \lambda_I &=& - \frac{1}{2} \gamma^{\mu\nu} \xi_I \mathcal{F}_{\mu\nu} + \gamma^\mu \xi_I \mathrm{D}_\mu \sigma - D_{IJ} \xi^J + \frac{\ii}{3\sqrt{2}} \gamma^{\mu\nu} \xi_I b_{\mu\nu} \sigma - \frac{2\sqrt{2}  \ii}{3} \tilde{\xi}_I \sigma ~, \nn \\
\delta_\xi D_{IJ}  &=& - 2 \ii \xi_{(I} \gamma^\mu \mathrm{D}_\mu \lambda_{J)} + 2 [ \sigma ,\xi_{(I} \lambda_{J)} ] + \frac{2\sqrt{2} }{3} \tilde{\xi}_{(I} \lambda_{J)} - \frac{1}{6 \sqrt{2}} \xi_{(I} \gamma^{\mu\nu} \lambda_{J)} b_{\mu\nu} \, . \label{VMsusys}
\eea
This has Grassmann odd supersymmetry parameters $\xi_I$, $\tilde{\xi}_I$.  We find that these transformations close onto
\bea
~[\delta_\xi,\delta_\eta]\sigma &=& -\ii v^\nu \mathrm{D}_\nu\sigma  - \frac{\sqrt{2} \ii}{3}\varrho\sigma~,  \\
~[\delta_\xi,\delta_\eta]\mathcal{A}_\mu &  = & -\ii v^\nu \mathcal{F}_{\nu\mu} + D_\mu\Upsilon~, \nonumber \\
~[\delta_\xi,\delta_\eta]\lambda_I &=&
-\ii v^\nu \mathrm{D}_\nu\lambda_I+\ii  [\Upsilon,\lambda_I]- \frac{ \sqrt{2} \ii}{3} \left[ \frac{3}{2}\varrho\lambda_I + {R}_I^{~J}\lambda_J - \frac{1}{4}{\Theta}^{\alpha\beta}\gamma_{\alpha\beta}\lambda_I \right]~, \nonumber \\
~[\delta_\xi,\delta_\eta]D_{IJ}& =&-\ii v^\nu \mathrm{D}_\nu D_{IJ}+\ii [\Upsilon, D_{IJ}]- \frac{ \sqrt{2} \ii}{3} \left[ 2\varrho D_{IJ}+{R}_I^{~K}D_{JK}+{R}_J^{~K}D_{IK} \right]~,\nn
\eea
where we have defined
\bea
v^\mu &=& \phantom{-} 2\varepsilon^{IJ}\xi_I\gamma^\mu\eta_J~, \nonumber \\
\Upsilon &=& -2\ii\varepsilon^{IJ}\xi_I\eta_J\sigma~, \nonumber \\
\varrho &=& -2\ii\varepsilon^{IJ}(\xi_I\tilde\eta_J-\eta_I\tilde\xi_J)~, \nonumber \\
{R}_{IJ} &=& -3\ii(\xi_I\tilde\eta_J+\xi_J\tilde\eta_I-\eta_I\tilde\xi_J-\eta_J\tilde\xi_I)~, \nonumber \\
{\Theta}^{\alpha\beta} &=&-2\ii\varepsilon^{IJ}(\tilde\xi_I\gamma^{\alpha\beta}\eta_J-\tilde\eta_I\gamma^{\alpha\beta}\xi_J) - {2\ii} \varepsilon^{IJ} ( \xi_I \eta_J ) b^{\alpha\beta} + \frac{\ii}{4 } \varepsilon^{\mu\nu\rho \alpha\beta} b_{\mu\nu} v_\rho ~~,
\eea
and ${R}_I^{~J}=\varepsilon^{JK} {R}_{IK}$, provided that the spinors $(\xi,\tilde{\xi})$ and $(\eta,\tilde{\eta})$ satisfy 
the $SU(2)_R$-covariantization of the $(\chi,\varphi)$ spinor equations (\ref{KSEchi})--(\ref{Dilchi}). 
More precisely
\begin{align}
\mathrm{D}_\mu \xi_I \ =& \ - \frac{\sqrt{2} \ii}{3} \gamma_\mu \tilde{\xi}_I - \frac{\ii}{12\sqrt{2}} b_{\nu\rho} \gamma_\mu{}^{\nu\rho} \xi_I + \frac{\ii}{3\sqrt{2}} b_{\mu\nu} \gamma^\nu \xi_I \, , \nn \\
\mathrm{D}_\mu \tilde{\xi}_I \ =& \ - \frac{\ii}{6\sqrt{2}} b_{\mu\nu} \gamma^\nu \tilde{\xi}_I - \frac{1}{16} \mathrm{D}^\nu b^{\rho\sigma} \gamma_{\mu\nu\rho\sigma} \xi_I + \frac{1}{16} \mathrm{D}_\mu b_{\nu\rho} \gamma^{\nu\rho} \xi_I - \frac{1}{8} \mathrm{D}_\nu b_{\mu\rho} \gamma^{\nu\rho} \xi_I \nn \\
&\ + \frac{\ii}{8\sqrt{2}} V_{\nu\rho I}{}^J \gamma_{\mu}{}^{\nu\rho} \xi_J - \frac{3\ii}{4\sqrt{2}} V_{\mu\nu I}{}^J \gamma^\nu \xi_J - \frac{1}{36} A^{(0)}_{\nu} \gamma_\mu{}^{\nu} \xi_I + \frac{1}{12} A^{(0)}_\mu \xi_I + \frac{\ii}{2} \omega_{\mu\nu} \gamma^\nu \xi_I \, , \nn \\
0 \ =& \ - \frac{1}{6\sqrt{2}} b_{\mu\nu} \gamma^{\mu\nu} \tilde{\xi}_I - \frac{\sqrt{2}}{3} X_2 \xi_I + \frac{\ii}{8} \mathrm{D}_\mu b_{\nu\rho} \gamma^{\mu\nu\rho} \xi_I - \frac{\ii}{18} A^{(0)}_\mu \gamma^\mu \xi_I - \frac{1}{4\sqrt{2}} V_{\mu\nu I}{}^J \gamma^{\mu\nu} \xi_J ~, \label{KSEsec4}
\end{align}
with $V_{\mu\nu}{}^{IJ}\equiv  2 \partial_{[\mu} V_{\nu]}{}^{IJ} - 2 V_{[\mu}{}^{K(I} V_{\nu] K}{}^{J)}$. 
Recall that $b$ has Weyl weight $w=1$, while the spinors have weight $w=\pm 1/2$.

It is crucial for the closure of the algebra that $\omega_{\mu\nu} = \omega_{\nu\mu}$, which is the same condition used in deriving the differential constraints (\ref{scalarbil})--(\ref{Omegabil}). 
Also as for that computation the closure of the supersymmetry algebra is 
 insensitive to the explicit form of $\omega_{\mu\nu}$, $A^{(0)}$ or $X_2$. Let us also notice that 
the supersymmetry variations (\ref{VMsusys}) reduce to those of the round $S^5$ in \cite{Hosomichi:2012ek} 
(in particular $b\equiv 0$ for the round $S^5$, and $\tilde{\xi}^{\mathrm{here}}_I=\frac{3}{\sqrt{2}}\ii \tilde{\xi}^{\mathrm{there}}_I$).

We now consider the on-shell hypermultiplet which consists of two complex scalars $q_I$ and a spinor $\psi$, all transforming in an arbitrary representation of the gauge group. A system of $r$ hypermultiplets is described by $q_I^A, \psi^A$ with $A=1, \ldots , 2r$. The fields satisfy the reality conditions $( q_I^A )^* = \Omega_{AB} \varepsilon^{IJ} q_J^B$ and $( \psi^{A\alpha} )^* = \Omega_{AB} \mathcal{C}_{\alpha\beta} \psi^{B\beta}$ with $\Omega_{AB}$ being the invariant tensor of $Sp(r)$. The supersymmetry variations for the system of $r$ hypermultiplets coupled to the vector multiplet are
\bea
\delta_\xi q_I^A &=&- 2 \ii \xi_I \psi^A \, , \nn \\
\delta_\xi \psi^A &=&\phantom{-} \varepsilon^{IJ} \gamma^\mu \xi_I \mathrm{D}_\mu q^A_J + \ii \varepsilon^{IJ} \xi_I \sigma q_J^A - \sqrt{2} \ii \varepsilon^{IJ} \tilde{\xi}_I q^A_J ~. \label{HMsusys}
\eea
The commutator of two supersymmetry transformations leads to
\begin{align}
~[ \delta_\xi , \delta_\eta ] q^A_I \ =& \ - \ii v^\mu \mathrm{D}_\mu q^A_I + \ii \Upsilon q_I^A - \frac{\sqrt{2}\ii}{3} \left[ \frac{3}{2} \varrho q^A_I + R_I{}^J q^A_J \right] \, , \nn \\
~[\delta_\xi,\delta_\eta] \psi^A \ =& \ - \ii v^\mu \mathrm{D}_\mu \psi^A + \ii \Upsilon \psi^A - \frac{\sqrt{2}\ii}{3} \left[ 2 \varrho \psi^A - \frac{1}{4} \Theta^{\alpha\beta} \gamma_{\alpha\beta} \psi^A \right] \nn \\
& \ + \frac{1}{2} v_\rho \Gamma^\rho \left( \ii \gamma^\mu \mathrm{D}_\mu \psi^A + \sigma \psi^A + \varepsilon^{IJ} \lambda_I q_J^A - \frac{1}{4\sqrt{2}} \gamma^{\mu\nu} \psi^A b_{\mu\nu} \right) \nn \\
& \ - \varepsilon^{KL} ( \xi_K \eta_L ) \left( \ii \gamma^\mu \mathrm{D}_\mu \psi^A + \sigma \psi^A + \varepsilon^{IJ} \lambda_I q_J^A - \frac{1}{4\sqrt{2}} \gamma^{\mu\nu} \psi^A b_{\mu\nu} \right) \, ,
\end{align}
where 
\bea
\mathrm{D}_\mu q_I^A &=& \partial_\mu q_I^A - \ii \mathcal{A}_\mu q_I^A - \frac{3}{2} \Cweyl_\mu q_I^A - V_{\mu I}{}^J q_J^A \nn \, , \\
\mathrm{D}_\mu \psi^A &=& \nabla_\mu \psi^A - \ii \mathcal{A}_\mu \psi^A - 2 \Cweyl_\mu \psi^A ~.
\eea
Closure of the algebra occurs only on-shell and this identifies the fermionic equation of motion as 
\bea
 E_\psi &\equiv& \ii \gamma^\mu \mathrm{D}_\mu \psi^A + \sigma \psi^A + \varepsilon^{IJ} \lambda_I q_J^A - \frac{1}{4\sqrt{2}} \gamma^{\mu\nu} \psi^A b_{\mu\nu}  \ = \ 0~. \label{HyperFermEOM}
\eea
Acting on $E_\psi$ with the supersymmetry transformations gives the bosonic equation of motion:
\begin{align}
 \varepsilon^{IJ} \bigg( \mathrm{D}^\mu \mathrm{D}_\mu q^A_J + \sigma^2 q_J^A - \frac{1}{3} X_2 q_J^A + \frac{1}{\sqrt{2}} \omega_{\mu}{}^\mu q^A_J - 2 ( \psi^A \lambda_J ) \bigg) + \ii D^{IJ} q_J^A \ = \ 0~. \label{HyperBosEOM}
\end{align}

\subsection{Lagrangians}

The action for a vector multiplet in five dimensions is determined by the prepotential $\mathscr{F}(\mathcal{V})$, which is a real and gauge invariant function of the vector superfield $\mathcal{V}$. Gauge invariance limits the prepotential to being at most cubic in $\mathcal{V}$ \cite{Seiberg:1996bd} and classically it takes the form
\bea
\mathscr{F}(\mathcal{V}) &=& \mathrm{Tr} \bigg[ \frac{1}{2 \mathtt{g}^2} \mathcal{V}^2 + \frac{k}{6} \mathcal{V}^3 \bigg] \, .\label{prepot}
\eea
Here $\mathtt{g}$ is the dimensionful gauge coupling constant and $k$ is a real constant which is subject to a quantization condition dependent on the gauge group \cite{Intriligator:1997pq}. Writing the components of the vector superfield as $\mathcal{V}^a T_a = ( \sigma^a T_a , \mathcal{A}_\mu^a T_a , \lambda_I^a T_a , D_{IJ}^a T_a) $ where $T_a$ are generators of the gauge group in the adjoint representation we find the cubic prepotential term in our curved backgrounds to be
\begin{align}
\mathcal{L}_{\mathrm{cubic}} \ =& \ \mathrm{d}_{abc} \bigg[  \frac{1}{24} \epsilon^{\mu\nu\rho\sigma\tau} \mathcal{A}^a_\mu \mathcal{F}^b_{\nu\rho} \mathcal{F}^c_{\sigma\tau} + \frac{\ii}{8} \varepsilon^{IJ} (\lambda_I^a \gamma^{\mu\nu} \lambda_J^b ) \mathcal{F}_{\mu\nu}^c + \frac{\ii}{4} D^{a, IJ} ( \lambda_I^b \lambda_J^c ) \bigg] \nn \\
&+ \mathrm{d}_{abc} \sigma^a \bigg[  \frac{1}{4} \mathcal{F}^b_{\mu\nu} \mathcal{F}^{c,\mu\nu} - \frac{1}{2} \mathrm{D}_\mu \sigma^b \mathrm{D}^\mu \sigma^c - \frac{1}{4} D_{IJ}^b D^{c,IJ} \nn \\
&\qquad \qquad \ - \frac{\ii}{2\sqrt{2}} \sigma^b \mathcal{F}^c_{\mu\nu} b^{\mu\nu} + \frac{1}{3} \sigma^b \sigma^c \left( \frac{\sqrt{2}}{3} \omega_\mu{}^\mu + \frac{2}{3} X_2 - \frac{5}{18} b_{\mu\nu} b^{\mu\nu} \right) \\
&\qquad \qquad \ + \frac{\ii}{2} \varepsilon^{IJ} ( \lambda_I^b \gamma^\mu \mathrm{D}_\mu \lambda_J^c ) - \frac{1}{2} \varepsilon^{IJ} \lambda_I^b [ \lambda_J, \sigma ]^c + \frac{1}{8\sqrt{2}} \varepsilon^{IJ} ( \lambda^b_I \gamma^{\mu\nu} \lambda^c_J ) b_{\mu\nu} \Bigg] ~. \nn
\end{align}
Here $\mathrm{d}_{abc} \propto \frac{k}{\pi^2} \mathrm{Tr} \left( T_{(a} T_b T_{c)} \right)$ is a symmetric invariant tensor of the gauge group. It vanishes\footnote{For example, taking the gauge group to be $G=SO(N)$ so that the Lie algebra generators satisfy $T_a^t = - T_a$ then $\mathrm{Tr} \left( T_{(a} T_b T_{c)} \right) = \mathrm{Tr} \left( T_{(a} T_b T_{c)} \right)^t = - \mathrm{Tr} \left( T_{(a} T_b T_{c)} \right)$.} for all simple gauge groups except $U(1)$ or $SU(N)$ with $N \geq 3$. The Lagrangian $\mathcal{L}_{\mathrm{cubic}}$ is invariant under the superconformal transformations \eqref{VMsusys} provided the supersymmetry parameters satisfy \eqref{KSEsec4}, and in addition $A^{(0)}$ is given by
\bea
A^{(0)} &=& - \frac{9}{4} * \left( ( \dd + 2 \Cweyl \wedge ) * b - \frac{\ii\sqrt{2}}{3} b \wedge b\right) \, ,
\eea
which matches precisely the expression (\ref{weylA0}) in section \ref{SecGeometry}.

The quadratic term in the prepotential includes Yang-Mills kinetic terms and is not conformally invariant. We therefore expect to break conformality by using the relation 
\bea
\tilde{\xi}_I &=& - \alpha_I{}^J \xi_J - \frac{\ii}{2} ( K_2 )_\mu \gamma^\mu \xi_I \, , \label{DilvarphiSU}
\eea
which is the $SU(2)_R$-covariantization of \eqref{Dilvarphi}. The Lagrangian describing the quadratic piece can be found from $\mathcal{L}_{\mathrm{cubic}}$ by identifying one of the vector superfields with a constant  supersymmetry preserving Abelian vector multiplet \cite{Imamura:2012xg}. That is
\bea
\mathcal{L}_{\mathrm{YM}} &=& \frac{1}{2\mathtt{g}^2} \mathcal{V}^{(1)} \mathrm{Tr} [ \mathcal{V}^2 ] \, ,
\eea 
where $\mathcal{V}^{(1)} = ( \sigma^{(1)} , \mathcal{A}_\mu^{(1)} , \lambda_I^{(1)} , D_{IJ}^{(1)} )$. We choose $\sigma^{(1)} =1$ and $\lambda_I^{(1)}=0$. Then $\mathcal{V}^{(1)}$ is supersymmetry preserving if the fermion variation 
\bea
\delta_\xi \lambda_I^{(1)} & =& - \frac{1}{2} \gamma^{\mu\nu} \xi_I \mathcal{F}^{(1)}_{\mu\nu} - D_{IJ}^{(1)} \xi^J + \frac{\ii}{3\sqrt{2}} \gamma^{\mu\nu} \xi_I b_{\mu\nu} - \frac{2\sqrt{2} \ii}{3} \alpha_{IJ} \xi^J - \frac{\sqrt{2}}{3} ( K_2 )_\mu \gamma^\mu \xi_I \, , \nn\\
&=& 0~,
\label{SUSYpreservingVM}
\eea
holds for non-trivial spinor parameters $\xi_I$ and some choice of $D^{(1)}_{IJ}$, $\mathcal{A}^{(1)}_\mu$ such that $\mathcal{F}^{(1)} = \dd \mathcal{A}^{(1)}$. Here we have substituted for $\tilde{\xi}_I$ using \eqref{DilvarphiSU}. To progress, note that there are two natural one-forms in our geometry namely $K_1$ and $K_2$. If we concentrate on $K_1$ which, with $S=1$, satisfies \eqref{K1bil}
\bea
 \dd K_1 & = &  \frac{2\sqrt{2}}{3} \left[ 2 \alpha J + K_1 \wedge K_2 + \ii  b - \frac{\ii}{2} i_{\xi} ( * b ) \right] \, ,
\eea
then upon $SU(2)_R$-covariantizing and  multiplying by $- \frac{1}{2} \gamma^{\mu\nu} \xi_I$ we  find
\bea
0 &=& - \frac{1}{2} \gamma^{\mu\nu} \xi_I \left( ( \dd K_1 )_{\mu\nu} - \frac{\ii\sqrt{2}}{3} b_{\mu\nu} \right) - \frac{\sqrt{2}}{3} \gamma^\mu \xi_I ( K_2 )_\mu - \frac{8\sqrt{2}\ii}{3} \alpha_{IJ} \xi^J \, .
\eea
To derive the previous equation we have used the projection conditions satisfied by the background geometry: $( K_1 )_\mu \gamma^\mu \chi = \chi$ and $J_{\mu\nu} \gamma^{\mu\nu} \chi = 4 \ii \chi$, along with $( K_1 )^\mu ( K_2 )_\mu =0 = ( K_1 )^\mu \Cweyl_\mu$ and $- \ii ( K_1 )^\mu b_{\mu\nu} = ( K_2 )_\nu + \tfrac{3}{\sqrt{2}} \Cweyl_\nu$. Comparing this to \eqref{SUSYpreservingVM} gives the constant vector multiplet as 
\bea
\mathcal{V}^{(1)} \ = \ ( \sigma^{(1)} , \mathcal{A}_\mu^{(1)} , \lambda_I^{(1)} , D_{IJ}^{(1)} ) \ = \ ( 1 ,(K_1)_\mu , 0 , 2\sqrt{2} \ii \alpha_{IJ} ) \, ,
\eea	
and the corresponding Yang-Mills Lagrangian is
\begin{align}
\mathcal{L}_{\mathrm{YM}} \ = \ \frac{1}{\mathtt{g}^2} \mathrm{Tr} \bigg[ & \frac{1}{4} \mathcal{F}_{\mu\nu} \mathcal{F}^{\mu\nu} - \frac{1}{2} \mathrm{D}_\mu \sigma \mathrm{D}^\mu \sigma - \frac{1}{4} D_{IJ} D^{IJ} + \frac{\ii}{2} \epsilon^{IJ} ( \lambda_I \gamma^\mu \mathrm{D}_\mu \lambda_J ) - \frac{1}{2} \varepsilon^{IJ} \lambda_I [ \lambda_J, \sigma ] \nn \\
&+ \frac{1}{8} \epsilon^{\mu\nu\rho\sigma\tau} \mathcal{F}_{\mu\nu} \mathcal{F}_{\rho\sigma} ( K_1 )_\tau - \frac{\ii}{\sqrt{2}} \sigma \mathcal{F}_{\mu\nu} b^{\mu\nu} + \frac{1}{2} \sigma \mathcal{F}^{\mu\nu} ( \dd K_1 )_{\mu\nu}  \label{FinalVMLagrangian} \\
& - 2\sqrt{2} \ii \sigma D^{IJ} \alpha_{IJ}+ \sigma^2 \left( \frac{\sqrt{2}}{3} \omega_\mu{}^\mu + \frac{2}{3} X_2 - \frac{5}{18} b_{\mu\nu} b^{\mu\nu} - \frac{\ii}{2\sqrt{2}} ( \dd K_1 )_{\mu\nu} b^{\mu\nu} \right)\nn \\
&+ \frac{\ii}{8} \varepsilon^{IJ} (\lambda_I \gamma^{\mu\nu} \lambda_J ) ( \dd K_1 )_{\mu\nu} + \frac{1}{8\sqrt{2}} \varepsilon^{IJ} ( \lambda_I \gamma^{\mu\nu} \lambda_J ) b_{\mu\nu} - \frac{1}{\sqrt{2}} ( \lambda_I \lambda_J ) \alpha^{IJ} \Bigg] \nn \, .
\end{align}
The second candidate one-form is $K_2$ but taking $\mathcal{F}^{(1)} = \dd K_2$ does not lead to \eqref{SUSYpreservingVM}.

The superconformal Lagrangian for the vector coupled hypermultiplets exists irrespective of the gauge group and is straightforward to construct: we simply integrate the equations of motion \eqref{HyperFermEOM} and \eqref{HyperBosEOM} found from closing the superalgebra to find 
\begin{align}
\mathcal{L}_{\mathrm{hm}} \ = \ \Omega_{AB} \bigg[ &- \frac{1}{2} \varepsilon^{IJ} \mathrm{D}^\mu q_I^A \mathrm{D}_\mu q_J^B + \frac{1}{2} \varepsilon^{IJ} q_I^A \sigma^2 q_J^B + \frac{\ii}{2} q_I^A D^{IJ} q_J^B \nn \\
&- 2 \varepsilon^{IJ} q_I^A ( \psi^B \lambda_J ) + \varepsilon^{IJ} q_I^A q_J^B \left( \frac{1}{2\sqrt{2}} \omega_{\mu}{}^\mu - \frac{1}{6} X_2 \right) \nn \\
&+ \ii ( \psi^A \gamma^\mu \mathrm{D}_\mu \psi^B ) + \psi^A \sigma \psi^B - \frac{1}{4\sqrt{2}} ( \psi^A \gamma^{\mu\nu} \psi^B ) b_{\mu\nu} \bigg] ~. \label{FinalHMLagrangian}
\end{align}
%

%%%%%%%%%%%%%%%%%%%%%%%%%%%%%%%%%%%%%%%%%%%%

\section{Discussion}\label{SecDiscussion}

In this paper we have constructed  rigid supersymmetric 
gauge theories with matter on a general class of five-manifold 
backgrounds. By construction these are the most general backgrounds 
that
arise as conformal boundaries 
of six-dimensional Romans supergravity solutions.
We find that $(M_5,g)$ is equipped with a conformal Killing 
vector which generates a transversely holomorphic foliation. 
In particular the transverse metric $g^{(4)}$ is an arbitrary Hermitian 
metric with respect to the transverse complex structure. 
This is a natural hybrid/generalization of the rigid 
supersymmetric geometries in three and four dimensions constructed 
in \cite{Closset:2012ru, Dumitrescu:2012ha, Klare:2013dka}, and 
includes  many previous constructions as special cases.

It is interesting to compare the geometry we find to the 
rigid limit of Poincar\'e supergravity \cite{Pan:2013uoa, Imamura:2014ima} 
and the twisting of \cite{Qiu:2014cha}. In the former case 
the backgrounds naively appear to be more general, as 
there is no almost complex structure singled out, nor 
integrability condition. However, they  don't include the $S^1\times S^4$ geometry relevant 
for the supersymmetric index, which as we showed in section \ref{SecExamples} 
\emph{is} included in our backgrounds.
In fact the singling out of the almost complex structure 
associated to $J=J_3$, where recall that $\Omega=J_2+\ii J_1$, in our geometry
is almost certainly related to the fact that in section \ref{SecGeometry} 
we focused on the case where we turn on only an Abelian $U(1)_R\subset SU(2)_R$. 
This was motivated in part for simplicity, and in part because the  known
solutions to Romans supergravity discussed previously also have this property.
Nevertheless, the supersymmetry variations and Lagrangians we constructed 
in section~\ref{SecLagrangians} are valid for an arbitrary background $SU(2)_R$ 
gauge field, and it should be relatively straightforward to analyse the geometric 
constraints in this more general case. Indeed, this is certainly necessary, and presumably sufficient,
to reproduce the partially  topologically twisted backgrounds $S^2\times M_3$ of \cite{Yagi:2013fda, Lee:2013ida, Cordova:2013cea}, since the $SU(2)$ spin connection of $M_3$ is twisted by $SU(2)_R$. 
On the other hand recall that the twisting in \cite{Qiu:2014cha} requires that $M_5$ be a 
K-contact manifold. This shares many features with our geometry, with 
one important difference: for a K-contact manifold the transverse 
two-form $J$ is closed, so the corresponding foliation is 
\emph{transversely symplectic}; however, our case is 
in some sense precisely the opposite, namely transversely holomorphic. 
These intersect precisely for Sasakian manifolds. 
It is interesting that these various approaches generally seem to 
lead to different supersymmetric geometries, with varying degrees of overlap.

Given the geometry we find and the results of \cite{Closset:2013vra}, it is natural to conjecture that the partition 
function and other BPS observables depend only on the transversely holomorphic 
foliation, {\it i.e.} for fixed such foliation they are independent of the 
choice of the remaining background data (functions $S$, $\alpha$, 
the one-form $\nu$ defined in section \ref{SecSummary}, and 
the transverse Hermitian metric $g^{(4)}$). It will be interesting 
to verify that this is indeed the case, and to compute these quantities 
using localization methods. Notice that \emph{locally} a transversely holomorphic 
foliation always looks like $\R\times \C^2$, which perhaps also explains 
why in \cite{Imamura:2014ima} the authors found that \emph{locally} 
all deformations of their backgrounds were $Q$-exact. 
Finally, our construction allows one to address holographic 
duals of these questions, which we plan to return to in future work.

%%%%%%%%%%%%%%%%%%%%%%%%%%%%%%%%%%%%%%%%%%%

\subsection*{Acknowledgments}

\noindent 
The work of L.~F.~A., M.~F. and P.~R. is supported by ERC STG grant 306260. L.~F.~A. is a Wolfson Royal Society Research Merit Award holder.  J.~F.~S. is supported by the Royal Society. P.~B.~G. is supported by EPSRC and a Scatcherd European Scholarship. 

%%%%%%%%%%%%%%%%%%%%%%%%%%%%%%%%%%%%%%%%%%%

\end{document}